\documentclass[12pt, draftclsnofoot, onecolumn]{IEEEtran}
\usepackage{amssymb}
\usepackage{amsmath}
\usepackage{cite}
\usepackage{url}
\usepackage{xcolor}
\usepackage{cite,graphicx,amsmath,amssymb}
\usepackage{subfigure}
\usepackage{citesort}
\usepackage{fancyhdr}
\usepackage{mdwmath}
\usepackage{mdwtab}
\usepackage{caption}
\usepackage{amsthm}
\usepackage{setspace}
\usepackage{algorithm}
\usepackage{algorithmic}
\usepackage{makecell}
\usepackage{diagbox}

\newtheorem{remark}{Remark}
\newtheorem{theorem}{Theorem}

\newtheorem{lemma}{Lemma}

\newtheorem{corollary}{Corollary}

\allowdisplaybreaks
\makeatletter
\def\ScaleIfNeeded{%
\ifdim\Gin@nat@width>\linewidth \linewidth \else \Gin@nat@width
\fi } \makeatother
\begin{document}
\title{Heterogeneous Semantic and Bit Communications: A Semi-NOMA Scheme}
\author{

Xidong~Mu,
        Yuanwei~Liu,~\IEEEmembership{Senior Member,~IEEE,}
       Li~Guo,~\IEEEmembership{Member,~IEEE,}
       and Naofal~Al-Dhahir,~\IEEEmembership{Fellow,~IEEE}
       
\thanks{Part of this work will be presented at the 18th International Symposium on Wireless Communication Systems (ISWCS), Hangzhou, China, October 19-22, 2022~\cite{Mu_ISWCS}.}
\thanks{Xidong Mu and Yuanwei Liu are with the School of Electronic Engineering and Computer Science, Queen Mary University of London, London E1 4NS, U.K. (e-mail: xidong.mu@qmul.ac.uk; yuanwei.liu@qmul.ac.uk).}
\thanks{Li Guo is with the Key Laboratory of Universal Wireless Communications, Ministry of Education, and the School of Artificial Intelligence, Beijing University of Posts and Telecommunications, Beijing 100876, China. (e-mail: guoli@bupt.edu.cn).}
\thanks{Naofal Al-Dhahir is with the Department of Electrical and Computer Engineering, The University of Texas at Dallas, Richardson, TX 75080 USA (e-mail: aldhahir@utdallas.edu).}
}

\maketitle
\vspace{-1.8cm}
\begin{abstract}
Multiple access (MA) design is investigated to facilitate the coexistence of the emerging semantic transmission and the conventional bit-based transmission in future networks. The \emph{semantic rate} is adopted for measuring the performance of the semantic transmission. However, a key challenge is that there is no closed-form expression for a key parameter, namely the \emph{semantic similarity}, which characterizes the sentence similarity between an original sentence and the corresponding recovered sentence. To overcome this challenge, we propose a data regression method, where the semantic similarity is approximated by a \emph{generalized logistic function}. Using the obtained tractable function, we propose a heterogeneous semantic and bit communication framework, where an access point simultaneously sends the semantic and bit streams to one semantics-interested user (S-user) and one bit-interested user (B-user). To realize this heterogeneous semantic and bit transmission in multi-user networks, three MA schemes are proposed, namely orthogonal multiple access (OMA), non-orthogonal multiple access (NOMA), and semi-NOMA. More specifically, the bit stream in semi-NOMA is split into two streams, one is transmitted with the semantic stream over the shared frequency sub-band and the other is transmitted over the separate orthogonal frequency sub-band. To study the fundamental performance limits of the three proposed MA schemes, the \emph{semantic-versus-bit (SvB) rate region} and the \emph{power region} are defined. An optimal resource allocation procedure is then derived for characterizing the boundary of the SvB rate region and the power region achieved by each MA scheme. The structures of the derived solutions demonstrate that semi-NOMA is superior to both NOMA and OMA given its highly flexible transmission policy. Our numerical results: 1) confirm that the proposed semi-NOMA is the optimal MA scheme as compared to OMA and NOMA even under the symmetric channel case, and 2) reveal that the superiority of semi-NOMA is more prominent when the channel condition of the S-user is better than that of the B-user.
\end{abstract}
\begin{IEEEkeywords}
Non-orthogonal multiple access, power region, semantic communications, semantic-versus-bit rate region.
\end{IEEEkeywords}

\section{Introduction}
Guided by Shannon's masterpiece published in 1948~\cite{Shannon1}, wireless communication systems have been developed rapidly from the first generation (1G) to the current fifth generation (5G) over the past few decades. Many efficient transmission technologies (e.g., massive multiple-input multiple-output (MIMO), millimeter wave communication, and advanced channel coding and modulation methods) have been proposed to approach the fundamental performance limits. Despite these achievements, the demand for capacity in wireless networks is expected to continue to grow explosively without limitations and for diverse and stringent scenarios, e.g., evolving from ``human-to-human'' communications to ``human-to-machine (H2M)'' and ``machine-to-machine (M2M)'' communications~\cite{ZP,H2M,M2M}. Facing these new challenges, researchers have begun to investigate the next-generation wireless network, namely the sixth-generation (6G). According to traditional reasoning, one way of achieving greater capacity is to increase the available bandwidth and number of antennas~\cite{6736761,8732419}. This, however, brings with it other challenges, such as extremely high energy consumption, hardware cost, and algorithmic complexity. To overcome these limitations, new technologies are needed for 6G.\\
\indent Recently, semantic communications have drawn significant attention from both industry and academia~\cite{Ping,Tong,Qin,Lan}. In contrast to conventional bit-based communication systems, the main idea of semantic communications is to transmit the semantic meaning contained in the source data. By doing so, the source data can be dramatically compressed and the required communication resources can be significantly reduced~\cite{Ping,Qin}. More importantly, semantic communications mesh well with envisioned H2M and M2M communications, where the main purpose is intelligent task execution, i.e., focusing on the semantic meaning instead of message distortion. It is worth mentioning that the concept of semantic communications is not a completely new idea. In Weaver and Shannon's book published in 1949~\cite{Shannon2}, three levels of communication problems were identified, namely a technical level, a semantic level, and an effectiveness level. Conventional bit-based communications have well solved the technical-level problem, i.e., \emph{How accurately can the symbols of communication be transmitted?} As a further advance, semantic communications respond to the semantic-level and effectiveness-level problems, i.e., \emph{How precisely do the transmitted symbols convey the desired meaning?} and \emph{How effectively does the received meaning affect conduct in the desired way?} However, given the previous technical limitations, semantic communications have not been systematically investigated. Fortunately, recent advancements of artificial intelligence and its effective applications in semantic information processing pave the way to develop semantic communications for future wireless networks.
\subsection{State-of-the-art}
In the past few decades, growing research efforts have been devoted to addressing the semantic-level and effectiveness-level communication problems defined in~\cite{Shannon2}. In contrast to the Shannon classical information theory employing the statistical probabilities of messages, the authors of \cite{Carnap} proposed the semantic information theory by employing the logical probabilities of the content of messages. Based on this, the authors of \cite{6004632} proposed a generic model for semantic communications, where the semantic noise and semantic channel capacity are defined for measuring semantic information. The authors of \cite{8476247} proposed a semantic communication framework, where the optimal transmission policy for minimizing the end-to-end average semantic error was derived using the Bayesian game. Apart from the theoretical study of semantic communications, researchers started to explore the implementation of semantic communications for delivering text, image, and video by employing powerful deep learning (DL) tools. For example, motivated by the success of employing DL in natural language processing (NLP), the authors of \cite{8461983} developed a DL-based joint source and channel coding (JSCC) approach for text transmission in the semantic domain. Moreover, the authors of \cite{8723589} further proposed a deep JSCC scheme for wireless image transmission, which does not employ explicit codes like conventional schemes and achieves superior performance in the low signal-to-noise ratio (SNR) and the limited channel bandwidth regimes. The authors of \cite{9464731} extended the deep JSCC into multi-channel image transmission and proposed a bandwidth-agile scheme for achieving high transmission performance. The authors of \cite{DeepSC_T} developed a DL-based joint semantic-channel coding method for text transmission, known as DeepSC, which is shown to outperform the deep JSCC proposed in \cite{8461983}. Furthermore, the DeepSC framework was extended by the authors of \cite{9450827} for speech transmission. Recently, based on the DeepSC text transmission~\cite{DeepSC_T}, the authors of \cite{SSE} proposed a new performance metric termed semantic rate, and studied the resource allocation problem in a semantic-aware multi-user communication network.
\subsection{Motivations and Contributions}
With the rapid breakthrough of the emerging semantic communications and the steady development of the currently employed bit-based communications, future networks are expected to simultaneously support both the semantic and bit transmissions to provide ubiquitous, customized, and intelligent connectivity among different types of devices (e.g., human, machine, and their interactions). However, since the available radio resources are limited, one of the most fundamental problems is how to design an efficient \emph{multiple access (MA)} scheme to facilitate the heterogeneous semantic and bit transmission for multi-user communications. Conventionally, MA schemes can be generally classified into orthogonal multiple access (OMA) and non-orthogonal multiple access (NOMA)~\cite{Liu2017}. It is well known that NOMA is capacity-achieving while OMA is suboptimal in the conventional bit-based transmission~\cite{Fundamentals}. Nevertheless, an interesting question arises, \emph{when considering semantic communications, does this conclusion still hold or do we need to develop new tailored MA schemes?}\\
\indent Driven by the above question, in this paper, we investigate the MA design for incorporating semantic communications into wireless networks. A heterogeneous semantic and bit communication framework is proposed, where an access point (AP) simultaneously sends the semantic and bit streams to one semantics-interested user (S-user) and one bit-interested user (B-user). To realize this challenging heterogeneous transmission under the given radio resources, on the one hand, the conventional OMA and NOMA schemes are proposed, where the two different streams are delivered via the orthogonal frequency sub-band and the fully shared frequency band, respectively. On the other hand, a novel MA scheme, namely semi-NOMA, is proposed, where part of the bit stream is delivered with the semantic stream via the shared frequency sub-band and the other part is delivered via the separate orthogonal frequency sub-band. Based on the three proposed MA schemes, two fundamental performance limits, namely the semantic-versus-bit (SvB) rate region and the power region, are characterized and compared to provide answers to the raised question. It suggests that \emph{NOMA does not necessarily always outperform OMA, while semi-NOMA is the optimal MA scheme in the new heterogeneous transmission.}\\
\indent The main contributions of this paper can be summarized as follows:
\begin{itemize}
  \item We exploit the semantic rate for characterizing the performance of the semantic transmission, which depends on the allocated frequency bandwidth and the achieved semantic similarity. However, the semantic rate is intractable due to the lack of a closed-form expression for the semantic similarity. To solve this problem, we propose to employ the data regression method and approximate the semantic similarity by a generalized logistic function with respect to the received SNR.
  \item We propose a heterogeneous semantic and bit transmission framework, where an AP simultaneously sends the semantic and bit streams to one S-user and one B-user. To facilitate this heterogeneous transmission, we propose three MA schemes, namely OMA, NOMA, and semi-NOMA. To study the fundamental performance limits of the three proposed MA schemes, we define 1) the SvB rate region, which consists of all the achievable semantic and bit rate-pairs, and 2) the power region, which determines the minimum transmit power to achieve the target semantic and bit rates.
  \item We first characterize the boundary of the SvB rate region achieved by the three proposed MA schemes via solving a series of resource allocation problems. We derive the optimal resource allocation policy for the three proposed MA schemes. In particular, the optimal power allocation among the two split bit streams in semi-NOMA follows the ``water-filling'' structure. By analyzing the structures of the derived solutions, it is revealed that the NOMA SvB rate region generally does not contain the OMA SvB rate region and both of them are contained in the semi-NOMA SvB rate region. Then, we characterize the power region of the three proposed MA schemes by optimally solving the resultant transmit power minimization problem.
  \item Our numerical results 1) validate our analysis and show that the proposed semi-NOMA is the optimal MA scheme in terms of both the SvB rate region and the power region. 2) Demonstrate that even if the channel gains of the S- and B-users are the same (i.e., the symmetric channel case), OMA is still strictly suboptimal as compared to semi-NOMA. 3) Reveal that pairing an S-user with a stronger channel gain with a B-user with a weaker channel gain together is preferable for employing semi-NOMA.
\end{itemize}
\subsection{Organization and Notation}
The rest of this paper is organized as follows: Section II introduces the new performance metric, namely the semantic rate, where the intractable semantic similarity component is well approximated by a generalized logistic function. Section III presents the system model, proposes three MA schemes to realize the heterogeneous semantic and bit transmission, and defines the SvB rate region and the power region. The SvB rate region and the power region of the three proposed MA schemes are characterized in Section IV and Section V, respectively. Section VI provides numerical examples and corresponding discussions. Finally, Section VII concludes the paper.\\
\indent \emph{Notation:} Scalars, vectors, and matrices are denoted by lower-case, bold-face lower-case, and bold-face upper-case letters, respectively. ${\mathbb{C}^{N \times 1}}$ denotes the space of $N \times 1$ complex-valued vectors. ${{\mathbf{a}}^H}$ denotes the conjugate transpose of vector ${\mathbf{a}}$.
\section{Semantic Rate and Approximation}
In this section, we first present a brief introduction on the new performance metric for semantic communications, namely the semantic rate~\cite{SSE}, which is based on the DeepSC~\cite{DeepSC_T}, a state-of-the-art semantic text transmission tool. Then, we propose to employ the generalized logistic function to approximate the key parameter, termed the semantic similarity, in the semantic rate to facilitate our theoretical investigation of the heterogeneous semantic and bit transmission in this work.
\subsection{Semantic Rate}
\begin{figure}[!ht]
  \centering
  \includegraphics[width=6in]{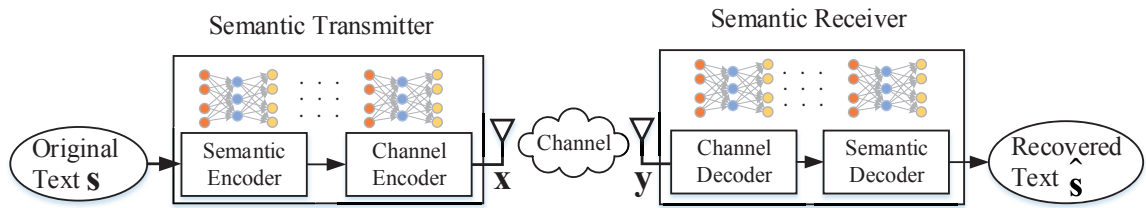}\\
  \caption{Illustration of the point-to-point DeepSC framework~\cite{DeepSC_T} for text transmission.}\label{P2P}
\end{figure}
\vspace{-0.5cm}
Let us take the typical point-to-point DeepSC text transmission framework~\cite{DeepSC_T} as an example, which is illustrated in Fig. \ref{P2P}. Let ${\mathbf{s}} = \left[ {{w_1},{w_2}, \ldots ,{w_L}} \right]$ denote an original sentence to be sent at the transmitter, where ${{w_l}}$ denotes the $l$th word in the sentence and $L$ denotes the average number of words per sentence. In contrast to conventional bit-based transceivers, the DeepSC transmitter (receiver) consists of two main components, namely semantic encoder (decoder) and channel encoder (decoder), which are empowered by well designed neural networks. The neural networks at the DeepSC transmitter are responsible for extracting semantic features and mapping them into semantic symbols to be transmitted, ${\mathbf{x}} \in {{\mathbb{C}}^{1 \times KL}}$, where $K \in {{\mathbb{Z}}^ + }$ denotes the average number of mapped semantic symbols for each word in ${\mathbf{s}}$. Considering the single-antenna DeepSC transmitter and receiver, the received signal at the DeepSC receiver can be expressed as ${\mathbf{y}} = h{\mathbf{x}} + {\mathbf{n}}$, where $h \in {{\mathbb{C}}^{1 \times 1}}$ and ${\mathbf{n}} \in {{\mathbb{C}}^{1 \times KL}}$ denote the complex wireless channel coefficient and the received noise symbol, respectively. The channel decoder and semantic decoder at the DeepSC receiver recover the original sentence, which is denoted by $\widehat {\mathbf{s}}$. In~\cite{DeepSC_T}, the performance of the DeepSC text transmission is evaluated by the semantic similarity between the original sentence, ${\mathbf{s}}$, and the corresponding recovered sentence, $\widehat {\mathbf{s}}$, which is given by
\begin{align}\label{sentence similarity}
\eta \left( {{\mathbf{s}},\widehat {\mathbf{s}}} \right) = \frac{{{\mathbf{B}}\left( {\mathbf{s}} \right){\mathbf{B}}{{\left( {\widehat {\mathbf{s}}} \right)}^T}}}{{\left\| {{\mathbf{B}}\left( {\mathbf{s}} \right)} \right\|\left\| {{\mathbf{B}}{{\left( {\widehat {\mathbf{s}}} \right)}}} \right\|}},
\end{align}
\vspace{-0.8cm}

\noindent where ${\mathbf{B}}\left( \cdot  \right)$ denotes the bidirectional encoder representations from transformers, a well-known model proposed in~\cite{peters-etal-2018-deep} for semantic information extraction in NLP. The above semantic similarity, $\eta \left( {{\mathbf{s}},\widehat {\mathbf{s}}} \right)$, ranges from 0 to 1 to indicate the similarity between ${\mathbf{s}}$ and $\widehat {\mathbf{s}}$ from a semantic perspective, where a higher value of $\eta \left( {{\mathbf{s}},\widehat {\mathbf{s}}} \right)$ means a higher similarity. According to~\cite{DeepSC_T}, $\eta \left( {{\mathbf{s}},\widehat {\mathbf{s}}} \right)$ mainly depends on the employed average number of semantic symbols per word, $K$, (i.e., the employed semantic encoding/decoding scheme) and the SNR of the received signal, $\gamma $. As a result, we can express the semantic similarity as a function of $K$ and $\gamma $, i.e., $\eta \left( {{\mathbf{s}},\widehat {\mathbf{s}}} \right) = \varepsilon \left( {K,\gamma } \right)$.  \\
\indent Based on the semantic similarity, a new performance metric, namely the semantic rate, was proposed in~\cite{SSE} for measuring the semantic information transmission rate achieved by the DeepSC~\cite{DeepSC_T}. Let $I$ (\emph{semantic units (suts)}) denote the average amount of semantic information contained in the sentence, ${\mathbf{s}}$. Therefore, the semantic information per semantic symbol is given by $\frac{I}{{KL}}$ (\emph{suts/symbol}). Recall the fact that the symbol rate is equal to the transmission bandwidth, which is denoted by $W$. Hence, the corresponding effective semantic rate (\emph{suts/s}) is given by~\cite{SSE}
\begin{align}\label{effective SSE}
S = \frac{{WI}}{{KL}}\varepsilon \left( {K,\gamma } \right).
\end{align}
According to~\cite{SSE}, the value of $\varepsilon \left( {K,\gamma } \right)$ under different $K$ and $\gamma$ can be obtained by running the DeepSC tool~\cite{DeepSC_T}. Part of the obtained results are shown in Fig. \ref{SS_result} (blue dots), where $K = 3,4,5,8,10,20$ symbols/word and $\gamma  = \left[ { - 10:1:20} \right]$ dB.
\vspace{-0.5cm}
\subsection{Semantic Rate Approximation}
\begin{figure}[!t]
  \centering
  \includegraphics[width=3in]{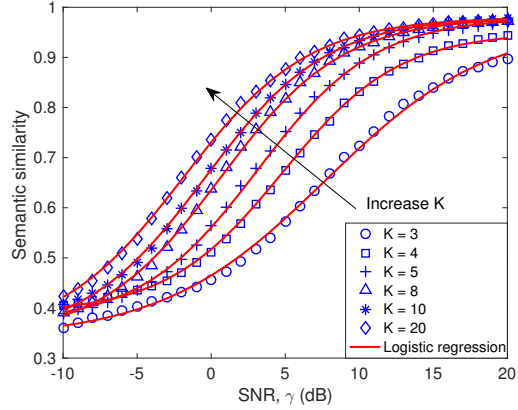}\\
  \caption{Generalized logistic regression for approximating the semantic similarity, $\varepsilon \left( {K,\gamma }\right)$.}\label{SS_result}
\end{figure}

Although semantic rate serves as an important performance metric for semantic communications, it is still non-trivial to theoretically investigate the corresponding semantic communication design since there is a lack of an explicit form for the semantic similarity, $\varepsilon \left( {K,\gamma } \right)$. To overcome this obstacle, in this work, we employ the data regression method to approximate the semantic similarity with respect to $\gamma$ for each $K$. By closely observing the presented results in Fig. \ref{SS_result} (blue dots), two insights can be summarized as follows. For any given $K$,\\
(i) $\varepsilon \left( {K,\gamma } \right)$ is monotonically non-decreasing with the increase of $\gamma$, and ${\varepsilon _{\min }} \le \varepsilon \left( {K,\gamma } \right) \le {\varepsilon _{\max }}$, where ${\varepsilon _{\min }},{\varepsilon _{\max }} \in \left[ {0,1} \right]$.\\
(ii) With increasing $\gamma$, the corresponding derivative, $\frac{d}{{d\gamma }}\varepsilon \left( {K,\gamma } \right)$, will in general first increase to a maximum value and then decrease.\\
\indent The above insights suggest that $\varepsilon \left( {K,\gamma } \right)$ should follow an `S' shape with respect to $\gamma$ and ranges from ${\varepsilon _{\min }}$ to ${\varepsilon _{\max }}$. This motivates us to employ the generalized logistic function to approximate $\varepsilon \left( {K,\gamma } \right)$, which is given by
\begin{align}\label{logistic}
\varepsilon \left( {K,\gamma } \right) \approx {\widetilde \varepsilon _K}\left( \gamma  \right) \triangleq {A_{K,1}} + \frac{{{A_{K,2}} - {A_{K,1}}}}{{1 + {e^{ - \left( {{C_{K,1}}\gamma  + {C_{K,2}}} \right)}}}}.
\end{align}
For any given $K$, ${A_{K,1}}>0$ and ${A_{K,2}}>0$ denote the lower (left) asymptote and the upper (right) asymptote, respectively, ${{C_{K,1}}}>0$ denotes the logistic growth rate, and ${{C_{K,2}}}$ controls the logistic mid-point. In this work, these parameters are determined by employing the minimum mean square error criterion for fitting the generalized logistic function to the data in Fig. \ref{SS_result} (blue dots). As it can be observed from Fig. \ref{SS_result}, for any given $K$, the constructed generalized logistic function (red lines) can achieve an accurate approximation\footnote{As assumed in~\cite{SSE,DeepSC_T}, $1 \le K \le 20$. For the clarity of Fig. \ref{SS_result}, we only present few values of $K$. It is worth noting that the proposed logistic regression is applicable for all values of $K$.}. More importantly, the approximated semantic similarity using the generalized logistic function in \eqref{logistic} has a tractable form, which can help to reveal fundamental design insights for semantic communications. For example, it can be observed from Fig. \ref{SS_result} that increasing $\gamma$ in the small $\gamma$ regime has a significant effect on the improvement of the semantic similarity and the enhancement vanishes in the high $\gamma$ regime, i.e., following the `S' shape. Moreover, it can also be observed that for a given $\gamma$, increasing $K$ in the small $K$ regime can achieve a more pronounced enhancement on the semantic similarity than that in the large $K$ regime.
\section{System Model}
\begin{figure}[!ht]
  \centering
  \includegraphics[width=3in]{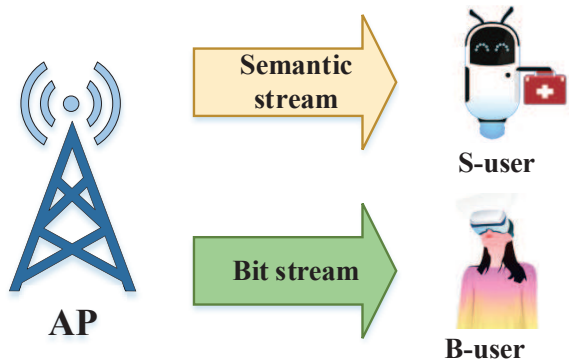}\\
  \caption{Illustration of the proposed heterogeneous semantic and bit communication framework.}\label{model}
\end{figure}
\vspace{-0.5cm}
As shown in Fig. \ref{model}, we propose a heterogeneous semantic and bit communication framework, where a single-antenna AP simultaneously transmits the semantic and bit streams to two single-antenna users. We refer to the user who is interested in the semantic stream as the S-user and the user who is interested in the bit stream as the B-user. In practice, the S-user can be intelligent robots in smart factories/hospitals, who are only interested in the meaning of text sent from the AP for task execution, and the B-user can be conventional human users. Let ${x_s}$ denote the normalized semantic symbol after the semantics-based encoder and ${x_b}$ denote the normalized information symbol after the conventional bit-based encoder. Therefore, the transmitted signal at the AP is given by
\vspace{-0.3cm}
\begin{align}\label{transmitted signal}
x = \sqrt {{p_s}} {x_s} + \sqrt {{p_b}} {x_b},
\end{align}
\vspace{-1.2cm}

\noindent where ${p_s} \ge 0$ and ${p_b} \ge 0$ denote the allocated transmit powers for the semantic and bit streams, respectively. Let $W$ and $P$ denote the total transmission frequency bandwidth and the maximum transmit power available at the AP. Given the limited radio resources, one fundamental issue is how to design an efficient MA scheme to facilitate this heterogeneous semantic and bit transmission. In the following, we will first propose two conventional MA schemes, namely OMA and NOMA, which are distinguished by whether the two streams are delivered via different orthogonal frequency sub-bands or the fully shared frequency band. Then, we will further propose a novel MA scheme, namely semi-NOMA, where the bit stream is split into two streams, one is delivered with the semantic stream via the shared frequency sub-band and the other is delivered via the separate orthogonal frequency sub-band. To study the fundamental performance limit of the heterogeneous communications, we assume that the channel state information (CSI) for all the users' channels can be perfectly obtained.
\subsection{OMA}
\begin{figure}[!ht]
  \centering
  \includegraphics[width=6.5in]{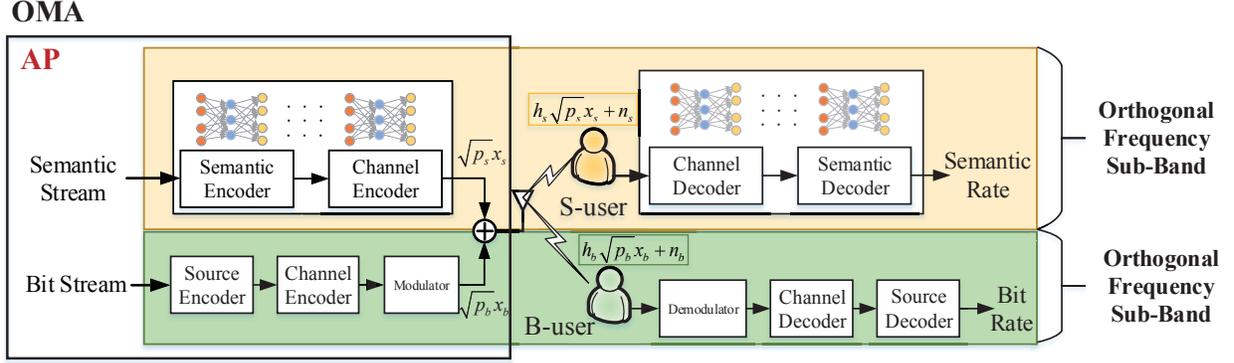}\\
  \caption{OMA based heterogeneous semantic and bit transmission, where the two streams are transmitted via two orthogonal frequency sub-bands.}\label{OMA}
\end{figure}
\vspace{-0.5cm}
In OMA, the AP simultaneously transmits the two streams to the S- and B-users via two orthogonal frequency bands, as illustrated in Fig. \ref{OMA}. Let ${W_s} \ge 0$ and ${W_b} \ge 0$ denote the allocated orthogonal frequency bandwidth for the semantic and bit streams, respectively, where ${W_s} + {W_b} = W$. Let ${h_s} \in {{\mathbb{C}}^{1 \times 1}}$ denote the channel coefficient from the AP to the S-user. As a result, the received SNR of the semantic signal at the S-user over $W_s$ is given by
\begin{align}\label{SNR_OMA}
{\gamma ^{\rm{O}}} = \frac{{{p_s}{{\left| {{h_s}} \right|}^2}}}{{{W_s}{N_0}}},
\end{align}
where ${{N_0}}$ is the received noise power spectral density. By employing the approximated semantic similarity metric proposed in \eqref{logistic}, the semantic rate achieved at the S-user in OMA can be expressed as follows:
\vspace{-0.2cm}
\begin{align}\label{SSE_OMA}
{S^{\rm{O}}} = \frac{{{W_s}I}}{{KL}}{\widetilde \varepsilon _K}\left( {{\gamma ^{\rm{O}}}} \right).
\end{align}
\vspace{-1cm}

\noindent Without loss of generality, in this paper, we define that ${S^{\rm{O}}} = \frac{{{W_s}I}}{{KL}}{\widetilde \varepsilon _K}\left( {\frac{{{p_s}{{\left| {{h_s}} \right|}^2}}}{{{W_s}{N_0}}}} \right) \triangleq 0$ when ${W_s} = 0$. \\
\indent Let ${h_b} \in {{\mathbb{C}}^{1 \times 1}}$ denote the channel coefficient from the AP to the B-user. The achievable communication rate of the bit signal follows the Shannon classical information theory, which is given by
\vspace{-0.3cm}
\begin{align}\label{CSE_OMA}
{R^{\rm{O}}} = {W_b}{\log _2}\left( {1 + \frac{{{p_b}{{\left| {{h_b}} \right|}^2}}}{{{W_b}{N_0}}}} \right).
\end{align}
\vspace{-1cm}

\indent It can be observed that the advantage of OMA is that the two streams are transmitted in an interference-free manner, which is easy to implement in practice. However, the disadvantage of OMA is that the spectrum efficiency would be limited.\\
\indent Given the available radio resources (i.e., $W$ and $P$) and the employed semantic encoding/decoding scheme, $K$, a fundamental performance trade-off between the semantic rate and the bit rate arises by employing the resource allocation among the two streams. To characterize this trade-off, we define the following SvB rate region:
\vspace{-0.3cm}
\begin{align}\label{Region_OMA}
{\mathcal{R}}_{{\rm{SvB}}}^{{\rm{O}}}\left( {W,P,K,\overline \varepsilon} \right) \triangleq \bigcup\limits_{{\mathbf{W}} \in {{\mathcal{F}}_1},{\mathbf{P}} \in {{\mathcal{F}}_2}} {\left\{ {\left( {S,R} \right):S \le {S^{\rm{O}}},{{\widetilde \varepsilon }_K}\left( {{\gamma ^{{\rm{O}}}}} \right) \ge \overline \varepsilon  ,R \le {R^{\rm{O}}}} \right\}},
\end{align}
\vspace{-0.8cm}

\noindent where ${\mathbf{W}} \triangleq \left\{ {{W_s},{W_b}} \right\}$ and ${\mathbf{P}} \triangleq \left\{ {{p_s},{p_b}} \right\}$ denote any feasible bandwidth and power allocation among the semantic and bit streams, respectively. The feasible set of ${\mathbf{W}}$ and ${\mathbf{P}}$ are, respectively, given by ${{\mathcal{F}}_1} \triangleq \left\{ {\left( {{W_s},{W_b}} \right):{W_s} + {W_b} = W,{W_s} \ge 0,{W_b} \ge 0} \right\}$ and ${{{\mathcal{F}}}_2} \triangleq \left\{ {\left( {{p_s},{p_b}} \right):{p_s} + {p_b} \le P,} \right.$\\$\left. {{p_s} \ge 0, {p_b} \ge 0} \right\}$. Here, ${\mathcal{R}}_{{\rm{SvB}}}^{{\rm{O}}}\left( {W,P,K,\overline \varepsilon} \right)$ consists of all the semantic and bit rate-pairs that can be achieved by OMA under the given $W$, $P$, $K$, and $\overline \varepsilon$. It is worth noting that compared to the bit rate, $R$, the performance of semantic transmission depends on both the semantic rate, $S$ and the achieved semantic similarity, ${{{\widetilde \varepsilon }_K}}$. This is because if the allocated bandwidth for the semantic stream, $W_s$, is sufficiently large, the resultant semantic rate can be large even if the achieved semantic similarity is quite low. Therefore, only considering the semantic rate may not guarantee the achieved semantic transmission performance. Hence, the achieved semantic similarity has to be also considered.\\
\indent Next, we define the \emph{power region} for OMA as follows:
\vspace{-0.3cm}
\begin{align}\label{PRegion_OMA}
P_{\min }^{{\rm{O}}}\left( {\overline S ,\overline \varepsilon  ,\overline R } \right) = \bigcap {\left\{ {P \in {{\mathbb{R}}^ + }:\left( {\overline S ,\overline R } \right) \in {\mathcal{R}}_{{\rm{SvB}}}^{{\rm{O}}}\left( {W,P,K,\overline \varepsilon  } \right)} \right\}},
\end{align}
\vspace{-1cm}

\noindent which determines the minimum required transmit power, $P_{\min }^O$, above which the target semantic rate, ${\overline S }$, semantic similarity, ${\overline \varepsilon  }$, and bit rate, ${\overline R }$, can be achieved by OMA under the given $W$ and $K$.
\subsection{NOMA}
\begin{figure}[h]
  \centering
  \includegraphics[width=6.5in]{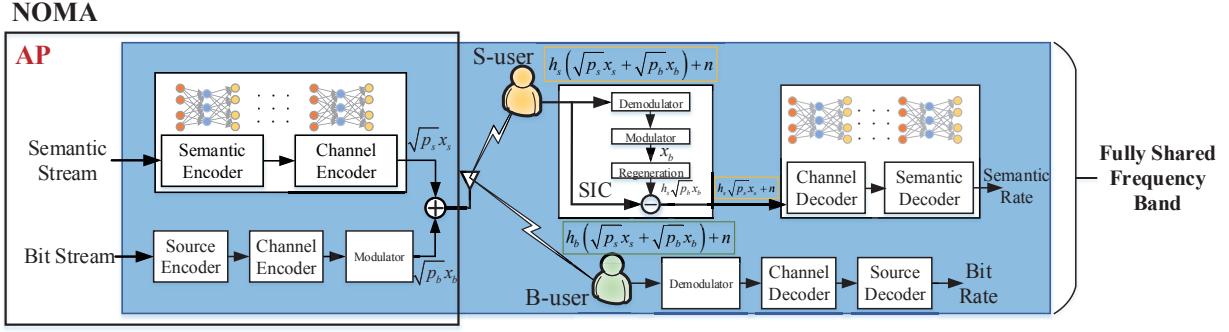}\\
  \caption{NOMA based heterogeneous semantic and bit transmission, where the two streams are transmitted via the fully shared frequency band.}\label{NOMA}
\end{figure}
In the conventional bit-based transmission, NOMA allows users to be served by sharing the same frequency band~\cite{Liu2017}. In particular, the transmitter sends the superimposed signal to users. Users having a better channel condition (i.e., higher channel gain) will employ successive interference cancellation (SIC) to first decode the signals of users having a worse channel condition (i.e., lower channel gain) and then subtract it from the received superimposed signal before decoding its own signal. However, in order to employ NOMA to facilitate the heterogeneous semantic and bit transmission, the decoding order of the two users cannot be determined by the users' channel conditions. The reasons can be explained as follows. On the one hand, the transmitter and receiver for semantic transmission have to be jointly trained in advance~\cite{DeepSC_T}, hence, it is impossible for the B-user to successfully decode the received semantic stream. On the other hand, even if the B-user was able to decode the semantic stream, it is also impossible to recover the semantic symbols for carrying out SIC as it is done in the conventional bit-based transmission. In contrast, employing the conventional bit-based encoding and decoding methods, the bit stream can be decoded and recovered at both the S-user\footnote{This requires the S-user to have the conventional bit-based decoder/encoder for carrying out SIC as shown in Fig. \ref{NOMA}, which is trivial to realize.} and B-user. Therefore, to facilitate the heterogeneous semantic and bit transmission with NOMA, we assume that a fixed bit-to-semantics decoding order is assumed, i.e., only the S-user is able to remove the B-user's signal via SIC but the B-user cannot.\\
\indent Motivated by this observation, as shown in Fig. \ref{NOMA}, the AP transmits the superimposed semantic and bit signal over the fully shared frequency band, $W$, in NOMA. At the S-user, the bit signal is firstly decoded from the received superimposed signal and the corresponding decoding rate is given by
\begin{align}\label{S_NOMA}
R_{b \to s}^{\rm{N}} = W{\log _2}\left( {1 + \frac{{{p_b}{{\left| {{h_s}} \right|}^2}}}{{{p_s}{{\left| {{h_s}} \right|}^2} + W{N_0}}}} \right).
\end{align}
To achieve the maximum performance gain of NOMA, perfect SIC is assumed in this paper. Therefore, after subtracting the bit signal from the received superimposed signal, the S-user will then decode its desired semantic signal and the achieved semantic rate is given by
\vspace{-0.2cm}
\begin{align}\label{SSE_NOMA}
{S^{\rm{N}}} = \frac{{WI}}{{KL}}{\widetilde \varepsilon _K}\left( {{\gamma ^{\rm{N}}}} \right),
\end{align}
\vspace{-1cm}

\noindent where ${\gamma ^{\rm{N}}} = \frac{{{p_s}{{\left| {{h_s}} \right|}^2}}}{{W{N_0}}}$.\\
\indent At the B-user, the bit signal will be directly decoded from the received superimposed signal by treating the semantic signal as interference. Then, the corresponding achievable bit rate is given by
\vspace{-0.2cm}
\begin{align}\label{W_NOMA}
R_{b \to b}^{\rm{N}} = W{\log _2}\left( {1 + \frac{{{p_b}{{\left| {{h_b}} \right|}^2}}}{{{p_s}{{\left| {{h_b}} \right|}^2} + W{N_0}}}} \right).
\end{align}
\vspace{-0.8cm}

\noindent Since the bit stream is decoded at both the S- and B-users, the effective communication rate of the bit stream is determined as follows~\cite{Liu2017}:
\vspace{-0.2cm}
\begin{align}\label{CSE_NOMA}
{R^{\rm{N}}} = \min \left\{ {R_{b \to s}^{\rm{N}},R_{b \to b}^{\rm{N}}} \right\} \triangleq W{\log _2}\left( {1 + \frac{{{p_b}{{\left| {{{\widetilde h}_b}} \right|}^2}}}{{{p_s}{{\left| {{{\widetilde h}_b}} \right|}^2} + W{N_0}}}} \right),
\end{align}
\vspace{-0.8cm}

\noindent where ${\left| {{{\widetilde h}_b}} \right|^2} \triangleq \min \left( {{{\left| {{h_s}} \right|}^2},{{\left| {{h_b}} \right|}^2}} \right)$ represents the effective channel gain for the bit transmission.\\
\indent Compared to OMA, NOMA may achieve a higher spectrum efficiency by transmitting the two streams via the fully shared frequency band and the semantic stream can also be delivered in an interference-free manner with the aid of SIC. However, using SIC causes the S-user to have a higher hardware complexity, as illustrated in Fig. \ref{NOMA}.\\
\indent Accordingly, the SvB rate region achieved by NOMA is defined as follows:
\vspace{-0.3cm}
\begin{align}\label{Region_NOMA}
{\mathcal{R}}_{{\rm{SvB}}}^{{\rm{N}}}\left( {W,P,K,\overline \varepsilon  } \right) \triangleq \bigcup\limits_{{\mathbf{P}} \in {{\mathcal{F}}_2}} {\left\{ {\left( {S,R} \right):S \le {S^{\rm{N}}},{{\widetilde \varepsilon }_K}\left( {{\gamma ^{{\rm{N}}}}} \right) \ge \overline \varepsilon  ,R \le {R^{\rm{N}}}} \right\}}.
\end{align}
\vspace{-0.8cm}

\noindent The corresponding power region of NOMA to achieve the target semantic and bit rates is given by
\vspace{-0.3cm}
\begin{align}\label{PRegion_NOMA}
P_{\min }^{{\rm{N}}}\left( {\overline S,\overline \varepsilon ,\overline R} \right) = \bigcap {\left\{ {P \in {\mathbb{R}^ + }:\left( {\overline S,\overline R} \right) \in {{\mathcal{R}}}_{{\rm{SvB}}}^{{\rm{N}}}\left( {W,P,K,\overline \varepsilon } \right)} \right\}}.
\end{align}
\subsection{Semi-NOMA}
\begin{figure}[h]
  \centering
  \includegraphics[width=6.5in]{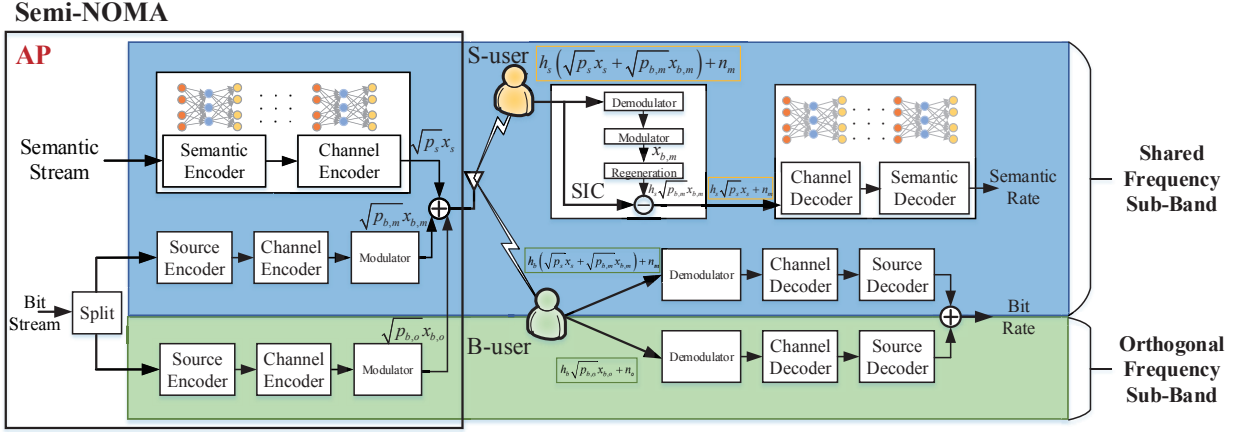}\\
  \caption{Semi-NOMA based heterogeneous semantic and bit transmission, where the original bit stream is split into two streams. One bit stream and the semantic stream are transmitted via the shared frequency sub-band, and the other bit stream is transmitted via the orthogonal frequency sub-band.}\label{SEMI}
\end{figure}
In this subsection, we propose a novel MA scheme, namely semi-NOMA, to achieve the heterogeneous semantic and bit transmission in a more flexible manner as compared to OMA and NOMA. As shown in Fig. \ref{SEMI}, the total frequency band is divided into two orthogonal frequency sub-bands. One is termed the semantic-bit shared frequency sub-band and the other is termed the bit-only orthogonal frequency sub-band. Accordingly, the original bit stream at the AP is split into two bit streams by allocating different powers. As illustrated in Fig. \ref{SEMI}, one bit stream is paired with the semantic stream following the NOMA principle, which are superimposed and transmitted to the two users over the shared frequency band. The other bit stream is transmitted to the B-user in an interference-free manner via the orthogonal frequency sub-band. Therefore, the transmission of the original bit stream is facilitated via the transmission of the two split bit streams over the two sub-bands. Let ${W_m}\ge 0$ and ${W_b}\ge 0$ denote the bandwidth of the shared and orthogonal frequency sub-band, where ${W_m}+{W_b}=W$. The transmitted signal at the AP is rewritten as follows:
\vspace{-0.3cm}
\begin{align}\label{transmitted signal1}
x = \sqrt {{p_s}} {x_s} + \sqrt {{p_{b,m}}} {x_{b,m}} + \sqrt {{p_{b,o}}} {x_{b,o}},
\end{align}
\vspace{-1cm}

\noindent where ${x_{b,m}}$ and ${x_{b,o}}$ denote the bit symbol in the shared and orthogonal frequency sub-bands, respectively, ${{p_{b,m}}} \le 0$ and ${{p_{b,o}}} \le 0$ represent the allocated transmit powers for the two bit streams, and ${p_s} + {p_{b,m}} + {p_{b,o}} \le P$.\\
\indent For the semantic and bit transmission over the shared frequency sub-band, it can be regarded like the previously-described NOMA scheme with a reduced bandwidth, ${W_m}$. Therefore, the decoding rate of the bit signal at the S-user over ${W_m}$ is given by
\vspace{-0.2cm}
\begin{align}\label{S M_NOMA}
R_{b,m \to s}^{\rm{S-N}} = {W_m}{\log _2}\left( {1 + \frac{{{p_{b,m}}{{\left| {{h_s}} \right|}^2}}}{{{p_s}{{\left| {{h_s}} \right|}^2} + {W_m}{N_0}}}} \right).
\end{align}
\vspace{-0.8cm}

\noindent After carrying out SIC, the achieved semantic rate at the S-user over ${W_m}$ is given by
\vspace{-0.2cm}
\begin{align}\label{SSE_S NOMA}
{S^{\rm{S-N}}} = \frac{{{W_m}I}}{{KL}}{\widetilde \varepsilon _K}\left( {{\gamma ^{\rm{S-N}}}} \right),
\end{align}
\vspace{-0.8cm}

\noindent where ${\gamma ^{\rm{S-N}}} = \frac{{{p_s}{{\left| {{h_s}} \right|}^2}}}{{{W_m}{N_0}}}$. Similarly, we define that ${S^{\rm{S-N}}}\left( {{W_m} = 0} \right) \triangleq 0$.\\
\indent The achievable communication rate of the bit signal at the B-user over ${W_m}$ is given by
\vspace{-0.2cm}
\begin{align}\label{CS M_NOMA}
R_{b,m \to b}^{\rm{S-N}} = {W_m}{\log _2}\left( {1 + \frac{{{p_{b,m}}{{\left| {{h_b}} \right|}^2}}}{{{p_s}{{\left| {{h_b}} \right|}^2} + {W_m}{N_0}}}} \right).
\end{align}
\vspace{-0.8cm}

\noindent Similarly, due to the employment of SIC, the effective communication rate of the bit signal over ${W_m}$ is given by~\cite{7445146}
\vspace{-0.2cm}
\begin{align}\label{CSE_M NOMA}
R_{b,m}^{\rm{S-N}} = \min \left\{ {R_{b,m \to s}^{\rm{S-N}},R_{b,m \to b}^{\rm{S-N}}} \right\} \triangleq {W_m}{\log _2}\left( {1 + \frac{{{p_{b,m}}{{\left| {{{\widetilde h}_b}} \right|}^2}}}{{{p_s}{{\left| {{{\widetilde h}_b}} \right|}^2} + {W_m}{N_0}}}} \right).
\end{align}
\vspace{-0.8cm}

\indent For the other bit signal transmitted over ${W_b}$, the corresponding achievable communication rate can be expressed as
\vspace{-0.2cm}
\begin{align}\label{CSE_B NOMA}
R_{b,o}^{\rm{S-N}} = {W_b}{\log _2}\left( {1 + \frac{{{p_{b,o}}{{\left| {{h_b}} \right|}^2}}}{{{W_b}{N_0}}}} \right).
\end{align}
\vspace{-0.8cm}

\noindent Therefore, the overall bit rate achieved by semi-NOMA is given by
\vspace{-0.2cm}
\begin{align}\label{CSE_S NOMA}
{R^{\rm{S-N}}} = R_{b,m}^{\rm{S-N}} + R_{b,o}^{\rm{S-N}}.
\end{align}
\vspace{-0.9cm}

\indent It can be observed that the proposed semi-NOMA scheme not only unifies both OMA and NOMA as special cases but also provides more flexible transmission options. To demonstrate it, if we set $W_b=0$, the proposed semi-NOMA scheme reduces to NOMA. If we set ${{p_{b,m}}}=0$, the proposed semi-NOMA scheme becomes OMA. Other transmission strategies can also be achieved by semi-NOMA with different resource allocation schemes. Therefore, the advantage of semi-NOMA is that it provides a high degree of flexibility for the heterogeneous semantic and bit transmission. However, as it can be observed from Fig. \ref{SEMI}, the proposed semi-NOMA entails a relatively high hardware complexity at both the transmitter and receiver.\\
\indent Similarly, the achieved SvB rate region of semi-NOMA is defined as follows:
\vspace{-0.2cm}
\begin{align}\label{Region_S NOMA}
{\mathcal{R}}_{{\rm{SvB}}}^{{\rm{S-N}}}\left( {W,P,K,\overline \varepsilon  } \right) \triangleq \bigcup\limits_{{{\mathbf{W}}^{{\rm{S-N}}}} \in {{\mathcal{F}}_3},{{\mathbf{P}}^{{\rm{S-N}}}} \in {{\mathcal{F}}_4}} {\left\{ {\left( {S,R} \right):S \le {S^{\rm{S-N}}},{{\widetilde \varepsilon }_K}\left( {{\gamma ^{{\rm{S-N}}}}} \right) \ge \overline \varepsilon  ,R \le {R^{\rm{S-N}}}} \right\}},
\end{align}
\vspace{-0.6cm}

\noindent where ${{\mathbf{W}}^{{\rm{S-N}}}} \triangleq \left\{ {{W_m},{W_b}} \right\}$ and ${{\mathbf{P}}^{{\rm{S-N}}}} \triangleq \left\{ {{p_s},{p_{b,m}},{p_{b,o}}} \right\}$ specify the bandwidth and power allocation in semi-NOMA, respectively, and ${{\mathcal{F}}_3} \triangleq \left\{ {\left( {{W_m},{W_b}} \right):{W_m} + {W_b} = W,{W_m} \ge 0,{W_b} \ge 0} \right\}$ and ${{\mathcal{F}}_4} \triangleq \left\{ {\left( {{p_s},{p_b}} \right):{p_s} + {p_{b,m}} + {p_{b,o}} \le P,{p_s} \ge 0,{p_{b,m}} \ge 0,{p_{b,o}} \ge 0} \right\}$ denote the corresponding feasible set.\\
\indent Similarly, the power region achieved by semi-NOMA is given by
\vspace{-0.2cm}
\begin{align}\label{PRegion_S NOMA}
P_{\min }^{{\rm{S-N}}}\left( {\overline S ,\overline \varepsilon  ,\overline R } \right) = \bigcap {\left\{ {P \in {{\mathbb{R}}^ + }:\left( {\overline S ,\overline R } \right) \in {\mathcal{R}}_{{\rm{SvB}}}^{{\rm{S-N}}}\left( {W,P,K,\overline \varepsilon  } \right)} \right\}}.
\end{align}
\vspace{-0.8cm}

\begin{figure}[!ht]
  \centering
  \includegraphics[width=5in]{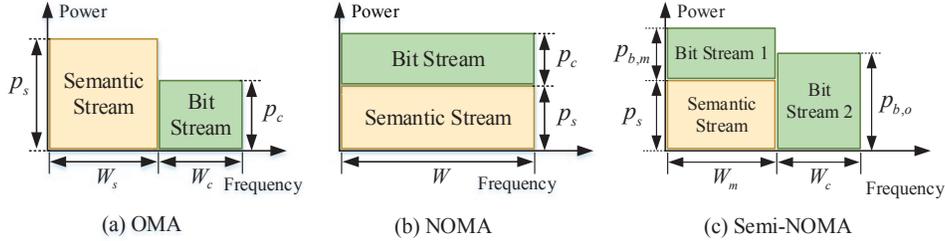}\\
  \caption{Illustration of the resource allocation features and variables in the three proposed MA schemes for the heterogeneous semantic and bit transmission.}\label{concept}
\end{figure}

\indent Fig. \ref{concept} illustrates the resource allocation features and variables of the three proposed MA schemes to realize the heterogeneous semantic and bit transmission.
\vspace{-0.5cm}
\section{SvB Rate Region Characterization}
For each proposed MA scheme, the optimal performance trade-off between the semantic and bit rate is characterized by the boundary of the corresponding SvB rate region. In this section, we will characterize all the boundary points in the SvB rate region for each proposed MA scheme.
\vspace{-0.5cm}
\subsection{SvB Rate Region Characterization for OMA}
For OMA, it is evident that there are two boundary points for indicating the extreme performance limits of semantic and bit transmission in ${\mathcal{R}}_{{\rm{SvB}}}^{{\rm{O}}}\left( {W,P,K,\overline \varepsilon} \right)$, which are denoted by $\left( {{S_{\max }},0} \right)$ and $\left( {0,{R_{\max }}} \right)$. We have
\vspace{-0.2cm}
\begin{align}\label{OMA S max}
{S_{\max }} = \frac{{WI}}{{KL}}{\widetilde \varepsilon _K}\left( {\frac{{P{{\left| {{h_s}} \right|}^2}}}{{W{N_0}}}} \right),
\end{align}
\vspace{-0.9cm}
\begin{align}\label{OMA R max}
{R_{\max }} = W{\log _2}\left( {1 + \frac{{P{{\left| {{h_b}} \right|}^2}}}{{W{N_0}}}} \right).
\end{align}
\vspace{-0.8cm}

\noindent Here, ${S_{\max }}$ is achieved by allocating all radio resources for realizing the semantic transmission, i.e., ${W_s} = W$ and ${p_s} = P$, and the resultant bit rate is zero. It is worth mentioning that for the SvB rate region characterization in this paper, we focus on the case of $\overline \varepsilon   \le {\widetilde \varepsilon _K}\left( {\frac{{P{{\left| {{h_s}} \right|}^2}}}{{W{N_0}}}} \right)$ to ensure that the boundary point $\left( {{S_{\max }},0} \right)$ is achievable in ${\mathcal{R}}_{{\rm{SvB}}}^{{\rm{O}}}\left( {W,P,K,\overline \varepsilon} \right)$. The proposed boundary characterization method is also applicable to the case of $\overline \varepsilon   > {\widetilde \varepsilon _K}\left( {\frac{{P{{\left| {{h_s}} \right|}^2}}}{{W{N_0}}}} \right)$. In this case, we can only allocate part of the frequency bandwidth (denoted by ${W_s^p}<W$) and all the transmit power to the semantic stream such that ${\widetilde \varepsilon _K}\left( {\frac{{P{{\left| {{h_s}} \right|}^2}}}{{W_s^p{N_0}}}} \right) = \overline \varepsilon$ to achieve the extreme boundary point $\left( {{{\widetilde S}_{\max }} = \frac{{W_s^pI}}{{KL}}\overline \varepsilon  ,0} \right)$. Since the corresponding SvB rate region will be reduced (i.e., ${\widetilde S_{\max }} < {S_{\max }}$) due to the insufficient transmit power, we term the case of $\overline \varepsilon   > {\widetilde \varepsilon _K}\left( {\frac{{P{{\left| {{h_s}} \right|}^2}}}{{W{N_0}}}} \right)$ as ``power-limited''. The case of $\overline \varepsilon   \le {\widetilde \varepsilon _K}\left( {\frac{{P{{\left| {{h_s}} \right|}^2}}}{{W{N_0}}}} \right)$ which we focus on is termed as ``power-sufficient''. Similarly, ${R_{\max }}$ is achieved by allocating all radio resources to the bit transmission, i.e., ${W_b} = W$ and ${p_b} = P$, and the resultant semantic rate is zero\footnote{It is worth noting that in this extreme case, we can think that the semantic similarity constraint can still be satisfied. This is because we can have $\mathop {\lim }\limits_{{W_s}\to 0,{p_s} \to 0} {\widetilde \varepsilon _K}\left( {\frac{{{p_s}{{\left| {{h_s}} \right|}^2}}}{{{W_s}{N_0}}}} \right) = \overline \varepsilon  $, while $\mathop {\lim }\limits_{{W_s}\to 0,{p_s} \to 0} \frac{{{W_s}I}}{{KL}}{\widetilde \varepsilon _K}\left( {\frac{{{p_s}{{\left| {{h_s}} \right|}^2}}}{{{W_s}{N_0}}}} \right) = 0$. Therefore, $\left( {0,{R_{\max }}} \right)$ can be regarded as one achievable boundary point in ${\mathcal{R}}_{{\rm{SvB}}}^{{\rm{O}}}\left( {W,P,K,\overline \varepsilon} \right)$.}. As a result, the remaining boundary points in ${\mathcal{R}}_{{\rm{SvB}}}^{{\rm{O}}}\left( {W,P,K,\overline \varepsilon} \right)$ that have to be characterized are over the intervals: $0 \le S \le {S_{\max }}$ and $0 \le R \le {R_{\max }}$. Then, we consider the following optimization problem:
\vspace{-0.3cm}
\begin{subequations}\label{OMA boundary}
\begin{align}
\mathop {\max }\limits_{\left\{ {{W_i},{p_i},i \in \left\{ {s,b} \right\}} \right\}} &\;{W_b}{\log _2}\left( {1 + \frac{{{p_b}{{\left| {{h_b}} \right|}^2}}}{{{W_b}{N_0}}}} \right)\\
\label{S OMA}{\rm{s.t.}}\;\;&\frac{{{W_s}I}}{{KL}}{\widetilde \varepsilon _K}\left( {\frac{{{p_s}{{\left| {{h_s}} \right|}^2}}}{{{W_s}{N_0}}}} \right) \ge \overline S ,\\
\label{SS OMA}&{\widetilde \varepsilon _K}\left( {\frac{{{p_s}{{\left| {{h_s}} \right|}^2}}}{{{W_s}{N_0}}}} \right) \ge \overline \varepsilon  ,\\
\label{Bandwidth allocation OMA}&{W_s} + {W_b} = W,{W_s} \ge 0,{W_b} \ge 0,\\
\label{power allocation OMA}&{p_s} + {p_b} \le P,{p_s} \ge 0,{p_b} \ge 0.
\end{align}
\end{subequations}
\vspace{-1cm}

\noindent where $\overline S $ represents the target semantic rate. By optimally solving problem \eqref{OMA boundary} with $0 \le \overline S \le {S_{\max }}$, the complete boundary of ${\mathcal{R}}_{{\rm{SvB}}}^{{\rm{O}}}\left( {W,P,K,\overline \varepsilon} \right)$ can be characterized. It can be verified that at the optimal solution for any given $0 \le \overline S \le {S_{\max }}$, it must hold that ${p_s} + {p_b} = P$. Therefore, problem \eqref{OMA boundary} can be rewritten as follows:
\begin{subequations}\label{OMA boundary 1}
\begin{align}
\mathop {\max }\limits_{{W_s},{p_s}} &\;\left( {W - {W_s}} \right){\log _2}\left( {1 + \frac{{\left( {P - {p_s}} \right){{\left| {{h_b}} \right|}^2}}}{{\left( {W - {W_s}} \right){N_0}}}} \right)\\
\label{S OMA 1}{\rm{s.t.}}\;\;&{\widetilde \varepsilon _K}\left( {\frac{{{p_s}{{\left| {{h_s}} \right|}^2}}}{{{W_s}{N_0}}}} \right) \ge \frac{{\overline S KL}}{{{W_s}I}},\\
\label{SS OMA 1}&{\widetilde \varepsilon _K}\left( {\frac{{{p_s}{{\left| {{h_s}} \right|}^2}}}{{{W_s}{N_0}}}} \right) \ge \overline \varepsilon  ,\\
\label{Allocation OMA}&0 \le {W_s} \le W,0 \le {p_s} \le P.
\end{align}
\end{subequations}
To solve problem \eqref{OMA boundary 1}, we have the following lemma.
\begin{lemma}\label{OMA2}
\emph{For the given $0 \le \overline S \le {S_{\max }}$, the optimal bandwidth allocation for the semantic transmission, $W_s^*$, must satisfy $W_s^* \in \left[ {{W^{low}},{W^{up}}} \right]$, where ${W^{low}} \triangleq \frac{{\overline S KL}}{I}$ and ${W^{up}} \triangleq \min \left( {\frac{{\overline SKL}}{{\overline \varepsilon I}},W} \right)$.}
\begin{proof}
\noindent Note that $0 \le {\widetilde \varepsilon _K} \le 1$, $\frac{{\overline S KL}}{{{W_s}I}} \le 1$ is necessary to guarantee the feasibility of problem \eqref{OMA boundary 1}, which leads to ${W_s^*} \ge \frac{{\overline S KL}}{I} \triangleq {W^{low}}$. Next, we determine the upper bound on $W_s^*$ using the fact that at the optimal solution to problem \eqref{OMA boundary 1}, the value of the right-hand-side of \eqref{S OMA 1} must be no less than $\overline \varepsilon $. This can be proved by contradiction. Suppose that the optimal solution $\left( {W_s^*,p_s^*} \right)$ to problem \eqref{OMA boundary 1} is achieved on the condition that $\frac{{\overline S KL}}{{W_s^*I}} < \overline \varepsilon $. Then, we can always construct a new feasible solution $\left( {\widehat W_s^*,\widehat p_s^*} \right)$, which achieves a higher objective value. To facilitate the design, we first construct $\left( {\widehat W_s^*,\widehat p_s^*} \right)$ such that $\frac{{\overline SKL}}{{\widehat W_s^*I}} = \overline \varepsilon $ and $\frac{{\widehat p_s^*}}{{\widehat W_s^*}} = \frac{{p_s^*}}{{W_s^*}}$. As the semantic similarity ${\widetilde \varepsilon _K}$ depends on the received SNR, $\left( {\widehat W_s^*,\widehat p_s^*} \right)$ is also feasible to problem \eqref{OMA boundary 1} to satisfy the constraint \eqref{SS OMA 1}, i.e., ${\widetilde \varepsilon _K}\left( {\frac{{\widehat p_s^*{{\left| {{h_b}} \right|}^2}}}{{\widehat W_s^*{N_0}}}} \right) \ge \overline \varepsilon$. Since $\left( {\widehat W_s^*,\widehat p_s^*} \right) \prec \left( {W_s^*,p_s^*} \right)$, the current objective value must be larger than that achieved by $\left( {W_s^*,p_s^*} \right)$. This suggests that the assumed $\left( {W_s^*,p_s^*} \right)$ satisfying $\frac{{\overline S KL}}{{W_s^*I}} < \overline \varepsilon $ cannot be the optimal solution to \eqref{OMA boundary 1}. Therefore, we have $\frac{{\overline SKL}}{{W_s^*I}} \ge \overline \varepsilon $, which yields $W_s^* \le \frac{{\overline SKL}}{{\overline \varepsilon I}}$. $W_s^*$ is also upper-bounded by the maximum frequency bandwidth, $W$. Therefore, $W_s^* \le \min \left( {\frac{{\overline SKL}}{{\overline \varepsilon I}},W} \right) \triangleq {W^{up}}$. The proof of Lemma \ref{OMA2} is thus completed.
\end{proof}
\end{lemma}
\indent The proof of \textbf{Lemma \ref{OMA2}} shows that the constraint \eqref{SS OMA 1} is superfluous and only \eqref{S OMA 1} needs to be considered when solving problem \eqref{OMA boundary 1}. It also implies that when the required semantic rate $\overline S$ is low, a smaller bandwidth should be allocated to the semantic stream for maximizing the bit rate. However, once $W_s$ is reduced such that $\frac{{\overline S KL}}{{W_sI}} > \overline \varepsilon $, it is generally unknown whether further reducing $W_s$ is beneficial for improving $R$ or not. This is because in this case, although reducing $W_s$ allows more bandwidth to be allocated to the bit transmission, more transmit power, $p_s$, may be required in the semantic transmission to achieve $\overline S$. Therefore, the objective value of problem \eqref{OMA boundary 1} is generally not monotonic with respect to $W_s$. To optimally solve problem \eqref{OMA boundary 1}, by employing the approximated function proposed in \eqref{logistic}, \eqref{S OMA 1} can be rewritten as follows:
\vspace{-0.2cm}
\begin{align}\label{p OMA}
{p_s} \ge \frac{{{W_s}{N_0}}}{{{{\left| {{h_s}} \right|}^2}}}{10^{\frac{{ - {C_{K,2}} - \ln \left( {\frac{{{A_{K,2}} - {A_{K,1}}}}{{\frac{{\overline SKL}}{{{W_s}I}} - {A_{K,1}}}}} -1\right)}}{{10{C_{K,1}}}}}} \triangleq p_{K}\left( {\overline S,{W_s}} \right).
\end{align}
\vspace{-0.8cm}

\noindent Note that for any given $W_s\in \left[ {{W^{low}},{W^{up}}} \right]$, if ${p_K}\left( {\overline S,{W_s}} \right)>P$, problem \eqref{OMA boundary 1} is infeasible and we define the corresponding objective value as zero for simplicity. Therefore, for any given $0 \le \overline S \le {S_{\max }}$, the optimal objective value of problem \eqref{OMA boundary 1} can be obtained as follows:
\vspace{-0.2cm}
\begin{align}\label{R OMA}
{R^*} = \mathop {\arg \max }\limits_{{W_s} \in \left[ {{W^{low}},{W^{up}}} \right]} \;\left( {W - {W_s}} \right){\log _2}\left( {1 + \frac{{\left[ {P - \overline {{p_K}\left( {\overline S,{W_s}} \right)} } \right]{{\left| {{h_b}} \right|}^2}}}{{\left( {W - {W_s}} \right){N_0}}}} \right),
\end{align}
\vspace{-0.6cm}

\noindent where $\overline {{p_K}\left( {\overline S,{W_s}} \right)}  \triangleq \min \left( {{p_K}\left( {\overline S,{W_s}} \right),P} \right)$. For any given $0 \le \overline S \le {S_{\max }}$, we only need to apply a one-dimensional search over ${{W_s} \in \left[ {{W^{low}},{W^{up}}} \right]}$ to obtain the corresponding optimal bit rate to problem \eqref{OMA boundary 1}. As a result, all the boundary points in ${\mathcal{R}}_{{\rm{SvB}}}^{{\rm{O}}}\left( {W,P,K,\overline \varepsilon} \right)$ can be characterized.
\subsection{SvB Rate Region Characterization for NOMA}
In this subsection, we will characterize the NOMA SvB rate region. By allocating all transmit power to the semantic transmission, $\left( {{S_{\max }},0} \right)$ is also one boundary point\footnote{In the power-limited case, ${\mathcal{R}}_{{\rm{SvB}}}^{{\rm{N}}}\left( {W,P,K,\overline \varepsilon} \right)= \emptyset$.} to achieve the extreme semantic performance limit in ${\mathcal{R}}_{{\rm{SvB}}}^{{\rm{N}}}\left( {W,P,K,\overline \varepsilon} \right)$. However, in contrast to OMA, where the allocated frequency bandwidth can be adjusted, the semantic and bit streams in NOMA always share the full frequency band. Due to the constraint on the semantic similarity, the semantic rate in ${\mathcal{R}}_{{\rm{SvB}}}^{{\rm{N}}}\left( {W,P,K,\overline \varepsilon} \right)$ is lower-bounded by
\vspace{-0.3cm}
\begin{align}\label{S min}
{S_{\min }^{\rm{N}}} = \frac{{WI}}{{KL}}\overline \varepsilon.
\end{align}
Let ${p_K}\left( {\overline \varepsilon  ,W} \right) \triangleq \frac{{W{N_0}}}{{{{\left| {{h_s}} \right|}^2}}}{10^{\frac{{ - {C_{K,2}} - \ln \left( {\frac{{{A_{K,2}} - {A_{K,1}}}}{{\overline \varepsilon   - {A_{K,1}}}}} -1\right)}}{{10{C_{K,1}}}}}}$ denote the minimum transmit power required to satisfy ${\widetilde \varepsilon _K}\left( {\frac{{{p_s}{{\left| {{h_s}} \right|}^2}}}{{W{N_0}}}} \right) \ge \overline \varepsilon$ in NOMA. In this case, by allocating the remaining transmit power to the bit transmission, the maximum bit rate in ${\mathcal{R}}_{{\rm{SvB}}}^{{\rm{N}}}\left( {W,P,K,\overline \varepsilon} \right)$ is given by
\vspace{-0.2cm}
\begin{align}\label{R max N}
R_{\max }^{\rm{N}} = W{\log _2}\left( {1 + \frac{{\left( {P - {p_K}\left( {\overline \varepsilon  ,W} \right)} \right){{\left| {{{\widetilde h}_b}} \right|}^2}}}{{{p_K}\left( {\overline \varepsilon  ,W} \right){{\left| {{{\widetilde h}_b}} \right|}^2} + W{N_0}}}} \right).
\end{align}
\vspace{-0.7cm}

\noindent Therefore, $\left( {{S_{\min }^{\rm{N}}},R_{\max }^{\rm{N}}} \right)$ is another extreme boundary point in ${\mathcal{R}}_{{\rm{SvB}}}^{{\rm{N}}}\left( {W,P,K,\overline \varepsilon} \right)$. Since only the transmit power can be adjusted in NOMA, all the boundary points between $\left( {{S_{\min }^{\rm{N}}},R_{\max }^{\rm{N}}} \right)$ and $\left( {{S_{\max }},0} \right)$ of ${\mathcal{R}}_{{\rm{SvB}}}^{{\rm{N}}}\left( {W,P,K,\overline \varepsilon} \right)$ can be characterized by sweeping $p_s$ from ${p_K}\left( {\overline \varepsilon  ,W} \right)$ to $P$. For each $p_s$, the corresponding boundary point is calculated using \eqref{SSE_NOMA} and \eqref{CSE_NOMA}, where ${p_b} = P - {p_s}$.
\begin{remark}\label{NOMA2}
\emph{It can be observed that the NOMA SvB rate region, ${\mathcal{R}}_{{\rm{SvB}}}^{{\rm{N}}}\left( {W,P,K,\overline \varepsilon} \right)$, does not necessarily contain the OMA SvB rate region, ${\mathcal{R}}_{{\rm{SvB}}}^{{\rm{O}}}\left( {W,P,K,\overline \varepsilon} \right)$. This is different from conventional bit-based communication, where NOMA is capacity-achieving and OMA is strictly suboptimal (except for the extreme points and the symmetric channel case)~\cite{Fundamentals}. This result answers the first part of the question raised in the introduction regarding the performance comparison between NOMA and OMA in the new heterogeneous semantic and bit transmission.}
\end{remark}
\vspace{-0.8cm}
\subsection{SvB Rate Region Characterization for semi-NOMA}
For semi-NOMA, $\left( {{S_{\max }},0} \right)$\footnote{In the power-limited case, this boundary point should be $\left( {{{\widetilde S}_{\max }},0} \right)$.} and $\left( {0,{R_{\max }}} \right)$ are also two extreme boundary points in the semi-NOMA SvB rate region, ${\mathcal{R}}_{{\rm{SvB}}}^{{\rm{S-N}}}\left( {W,P,K,\overline \varepsilon} \right)$. Similarly, the remaining boundary points of ${\mathcal{R}}_{{\rm{SvB}}}^{{\rm{S-N}}}\left( {W,P,K,\overline \varepsilon} \right)$ can be characterized by solving the following optimization problem for all $0 \le \overline S  \le {S_{\max }}$:
\vspace{-0.2cm}
\begin{subequations}\label{Semi boundary}
\begin{align}
\mathop {\max }\limits_{{W_m},{W_b},{p_s},{p_{b,m}},{p_{b,o}}} &\;{W_m}{\log _2}\left( {1 + \frac{{{p_{b,m}}{{\left| {{{\widetilde h}_b}} \right|}^2}}}{{{p_s}{{\left| {{{\widetilde h}_b}} \right|}^2} + {W_m}{N_0}}}} \right) + {W_b}{\log _2}\left( {1 + \frac{{{p_{b,o}}{{\left| {{h_b}} \right|}^2}}}{{{W_b}{N_0}}}} \right)\\
\label{S Semi}{\rm{s.t.}}\;\;&\frac{{{W_m}I}}{{KL}}{\widetilde \varepsilon _K}\left( {\frac{{{p_s}{{\left| {{h_s}} \right|}^2}}}{{{W_m}{N_0}}}} \right) \ge \overline S ,\\
\label{SS Semi}&{\widetilde \varepsilon _K}\left( {\frac{{{p_s}{{\left| {{h_s}} \right|}^2}}}{{{W_m}{N_0}}}} \right) \ge \overline \varepsilon  ,\\
\label{Bandwidth allocation Semi}&{W_m} + {W_b} = W,{W_m} \ge 0,{W_b} \ge 0,\\
\label{power allocation Semi}&{p_s} + {p_{b,m}} + {p_{b,o}} \le P,{p_s} \ge 0,{p_{b,m}} \ge 0, {p_{b,o}}\ge 0.
\end{align}
\end{subequations}
\vspace{-1.2cm}

\noindent For any given bandwidth allocation, $\left( {{W_m},{W_b}} \right)$, and target semantic rate, $\overline S$, by employing the approximated function proposed in \eqref{logistic}, constraints \eqref{S Semi} and \eqref{SS Semi} can be, respectively, rewritten as follows:
\vspace{-0.4cm}
\begin{subequations}
\begin{align}\label{C1}
{p_s} \ge& \frac{{{W_m}{N_0}}}{{{{\left| {{h_s}} \right|}^2}}}{10^{\frac{{ - {C_{K,2}} - \ln \left( {\frac{{{A_{K,2}} - {A_{K,1}}}}{{\frac{{\overline SKL}}{{{W_m}I}} - {A_{K,1}}}} - 1} \right)}}{{10{C_{K,1}}}}}} \triangleq {p_K}\left( {\overline S,{W_m}} \right),\\
\label{C2}{p_s} \ge& \frac{{{W_m}{N_0}}}{{{{\left| {{h_s}} \right|}^2}}}{10^{\frac{{ - {C_{K,2}} - \ln \left( {\frac{{{A_{K,2}} - {A_{K,1}}}}{{\overline \varepsilon   - {A_{K,1}}}} - 1} \right)}}{{10{C_{K,1}}}}}} \triangleq {p_K}\left( {\overline \varepsilon  ,{W_m}} \right),
\end{align}
\end{subequations}
\vspace{-1cm}

\noindent which are equivalent to the following constraint:
\vspace{-0.2cm}
\begin{align}\label{C3}
{p_s} \ge \max \left( {{p_K}\left( {\overline S,{W_m}} \right),{p_K}\left( {\overline \varepsilon  ,{W_m}} \right)} \right) \triangleq {p_K}\left( {\overline S,\overline \varepsilon  ,{W_m}} \right).
\end{align}
\vspace{-1cm}

\noindent Note that if ${p_K}\left( {\overline S,\overline \varepsilon  ,{W_m}} \right)>P$, problem \eqref{Semi boundary} is infeasible for the given bandwidth allocation and we define the corresponding objective value as zero for simplicity. The optimal solution to problem \eqref{Semi boundary} is achieved when the constraint \eqref{C3} becomes an equality, otherwise, we can always decrease $p_s$ to achieve a larger objective value. As a result, for any given ${W_m}$, problem \eqref{Semi boundary} is reduced to
\vspace{-0.2cm}
\begin{subequations}\label{Semi boundary 2}
\begin{align}
\mathop {\max }\limits_{{p_{b,m}},{p_{b,o}}} &\;{W_m}{\log _2}\left( {1 + \frac{{{p_{b,m}}{{\left| {{{\widetilde h}_b}} \right|}^2}}}{{{{p_K}\left( {\overline S,\overline \varepsilon  ,{W_m}} \right)}{{\left| {{{\widetilde h}_b}} \right|}^2} + {W_m}{N_0}}}} \right) + \left( {W - {W_m}} \right){\log _2}\left( {1 + \frac{{{p_{b,o}}{{\left| {{h_b}} \right|}^2}}}{{\left( {W - {W_m}} \right){N_0}}}} \right)\\
\label{power allocation Semi 2}{\rm{s.t.}}\;\;&{p_{b,m}} + {p_{b,o}} \le \widetilde P,{p_{b,m}} \ge 0, {p_{b,o}}\ge 0,
\end{align}
\end{subequations}
\vspace{-0.9cm}

\noindent where $\widetilde P \triangleq \max \left( {P - {p_K}\left( {\overline S,\overline \varepsilon  ,{W_m}} \right),0} \right)$. In this case, the objective bit rate can be maximized by the ``water-filling'' based power allocation~\cite{Fundamentals} subject to the power constraint $\widetilde P$. The optimal power allocations over the shared and orthogonal frequency sub-bands are given by
\vspace{-0.2cm}
\begin{align}\label{WF}
p_{b,m}^* = \left[ {\frac{{{W_m}}}{{{\lambda ^*}}} - \frac{1}{{{h_m}}}} \right]_0^{\widetilde P},p_{b,o}^* = \left[ {\frac{{{\left( {W - {W_m}} \right)}}}{{{\lambda ^*}}} - \frac{1}{{{h_b}}}} \right]_0^{\widetilde P},
\end{align}
\vspace{-0.6cm}

\noindent where $\left[ x \right]_a^b = \max \left( {\min \left( {x,b} \right),a} \right)$, ${h_m} \triangleq \frac{{{{\left| {{{\widetilde h}_b}} \right|}^2}}}{{{{p_K}\left( {\overline S,\overline \varepsilon  ,{W_m}} \right)}{{\left| {{{\widetilde h}_b}} \right|}^2} + {W_m}{N_0}}}$, ${h_b} \triangleq \frac{{{{\left| {{h_b}} \right|}^2}}}{{{\left( {W - {W_m}} \right)}{N_0}}}$, and ${\lambda ^*} \triangleq \frac{W}{{\widetilde P + \frac{1}{{{h_m}}} + \frac{1}{{{h_b}}}}}$. Let $R\left( {p_{b,m}^*,p_{b,o}^*|{W_m},\overline S } \right)$ denote the optimal objective value of problem \eqref{Semi boundary 2} under the given $W_m$ and $\overline S$. Therefore, for any given $0 \le \overline S  \le {S_{\max }}$, the optimal bit rate of problem \eqref{Semi boundary} can be obtained by employing a one-dimensional search over ${{W_m} \in \left[ {0,W} \right]}$ and the optimal water-filling power allocation \eqref{WF} under each given ${W_m}$, i.e.,
\vspace{-0.2cm}
\begin{align}\label{WF2}
{R^*} = \mathop {\arg \max }\limits_{{W_m} \in \left[ {0,W} \right]} \;R\left( {p_{b,m}^*,p_{b,o}^*|{W_m},\overline S } \right).
\end{align}
\vspace{-0.9cm}

\noindent As a result, all the boundary points between $\left( {{S_{\max }},0} \right)$ and $\left( {0,{R_{\max }}} \right)$ in ${\mathcal{R}}_{{\rm{SvB}}}^{{\rm{S-N}}}\left( {W,P,K,\overline \varepsilon} \right)$ can be completely characterized.
\begin{remark}\label{S-NOMA}
\emph{Recall the fact that both OMA and NOMA can be regarded as special cases of semi-NOMA, the semi-NOMA SvB rate region will always contain those achieved by OMA and NOMA, i.e., ${\mathcal{R}}_{{\rm{SvB}}}^{{\rm{S-N}}}\left( {W,P,K,\overline \varepsilon } \right) \supseteq {\mathcal{R}}_{{\rm{SvB}}}^{{\rm{N}}}\left( {W,P,K,\overline \varepsilon } \right)$ and ${\mathcal{R}}_{{\rm{SvB}}}^{{\rm{S-N}}}\left( {W,P,K,\overline \varepsilon } \right) \supseteq {\mathcal{R}}_{{\rm{SvB}}}^{{\rm{O}}}\left( {W,P,K,\overline \varepsilon } \right)$. This result answers the second part of the question raised in the introduction, which underscores the importance of developing a new MA scheme in the heterogeneous semantic and bit transmission. The superiority of semi-NOMA over OMA and NOMA will be further demonstrated by the numerical results provided in Section VI.}
\end{remark}
\section{Power Region Characterization}
In this section, we continue to characterize the power region achieved by each proposed MA scheme, which is defined in \eqref{PRegion_OMA}, \eqref{PRegion_NOMA}, and \eqref{PRegion_S NOMA}.
\subsection{Power Region Characterization for OMA}
The characterization of the power region is equivalent to minimizing the transmit power, subject to the constraints on the required semantic rate, $\overline S$, semantic similarity, $\overline \varepsilon$, and bit rate, $\overline R$. Then, we consider the following optimization problem for OMA:
\vspace{-0.4cm}
\begin{subequations}\label{OMA P}
\begin{align}
\mathop {\min }\limits_{{0 \le {W_s} \le W},{p_s\ge0},{p_b\ge0}} &\;p_s+p_b\\
\label{S O P}{\rm{s.t.}}\;\;&\frac{{{W_s}I}}{{KL}}{\widetilde \varepsilon _K}\left( {\frac{{{p_s}{{\left| {{h_s}} \right|}^2}}}{{{W_s}{N_0}}}} \right) \ge \overline S ,\\
\label{SS O P}&{\widetilde \varepsilon _K}\left( {\frac{{{p_s}{{\left| {{h_s}} \right|}^2}}}{{{W_s}{N_0}}}} \right) \ge \overline \varepsilon  ,\\
\label{Rate O P}&\left( {W - {W_s}} \right){\log _2}\left( {1 + \frac{{{p_b}{{\left| {{h_b}} \right|}^2}}}{{\left( {W - {W_s}} \right){N_0}}}} \right) \ge \overline R.
\end{align}
\end{subequations}
\vspace{-0.8cm}

\noindent It can be also verified that the optimal bandwidth allocation to problem \eqref{OMA P} satisfies $W_s^* \in \left[ {{W^{low}},{W^{up}}} \right]$. The proof is similar to \textbf{Lemma \ref{OMA2}}, and we omit it for brevity. Therefore, for any given $W_s \in \left[ {{W^{low}},{W^{up}}} \right]$, the minimum transmit power to \eqref{OMA P} is given by
\vspace{-0.2cm}
\begin{align}
P_{\min }^{\rm{O}}\left( {\overline S,\overline \varepsilon  ,\overline R |{W_s}} \right) = \frac{{\left( {W - {W_s}} \right){N_0}}}{{{{\left| {{h_b}} \right|}^2}}}\left( {{2^{\frac{{\overline R }}{{\left( {W -{W_s}} \right)}}}} - 1} \right) + {p_K}\left( {\overline S,{W_s}} \right),
\end{align}
\vspace{-0.8cm}

\noindent where ${p_K}\left( {\overline S,{W_s}} \right)$ is defined in \eqref{p OMA}. Accordingly, by employing a one-dimensional search over $W_s \in \left[ {{W^{low}},{W^{up}}} \right]$, the optimal solution to problem \eqref{OMA P} can be obtained as follows:
\vspace{-0.2cm}
\begin{align}
P_{\min }^{\rm{O}}\left( {\overline S,\overline \varepsilon  ,\overline R } \right) = \mathop {\arg \min }\limits_{{W_s} \in \left[ {{W^{low}},{W^{up}}} \right]} P_{\min }^{\rm{O}}\left( {\overline S,\overline \varepsilon  ,\overline R |{W_s}} \right).
\end{align}
\vspace{-1cm}
\subsection{Power Region Characterization for NOMA}
For NOMA, the resultant optimization problem for characterizing the power region can be written as follows:
\vspace{-0.3cm}
\begin{subequations}\label{NOMA P}
\begin{align}
\mathop {\min }\limits_{{p_s\ge0},{p_b\ge0}} &\;p_s+p_b\\
\label{S N P}{\rm{s.t.}}\;\;&\frac{{{W}I}}{{KL}}{\widetilde \varepsilon _K}\left( {\frac{{{p_s}{{\left| {{h_s}} \right|}^2}}}{{{W}{N_0}}}} \right) \ge \overline S ,\\
\label{SS N P}&{\widetilde \varepsilon _K}\left( {\frac{{{p_s}{{\left| {{h_s}} \right|}^2}}}{{{W}{N_0}}}} \right) \ge \overline \varepsilon  ,\\
\label{Rate N P}&W{\log _2}\left( {1 + \frac{{{p_b}{{\left| {{{\widetilde h}_b}} \right|}^2}}}{{{p_s}{{\left| {{{\widetilde h}_b}} \right|}^2} + W{N_0}}}} \right) \ge \overline R.
\end{align}
\end{subequations}
\vspace{-0.8cm}

\noindent By employing the approximated function proposed in \eqref{logistic}, constraints \eqref{S N P} and \eqref{SS N P} can be expressed as follows:
\vspace{-0.3cm}
\begin{align}
{p_s} \ge \max \left( {{p_K}\left( {\overline S,W} \right),{p_K}\left( {\overline \varepsilon ,W} \right)} \right) \triangleq {p_K}\left( {\overline S,\overline \varepsilon ,W} \right),
\end{align}
\vspace{-1cm}

\noindent where ${{p_K}\left( {\overline S,W} \right)}$ and ${{p_K}\left( {\overline \varepsilon ,W} \right)}$ are, respectively, obtained by replacing $W_m$ with $W$ in \eqref{C1} and \eqref{C2}. As a result, the optimal solution to problem \eqref{NOMA P} is given by
\vspace{-0.2cm}
\begin{align}
P_{\min }^{\rm{N}}\left( {\overline S,\overline \varepsilon  ,\overline R } \right) ={p_K}\left( {\overline S,\overline \varepsilon ,W} \right) + \frac{{{p_K}\left( {\overline S,\overline \varepsilon ,W} \right){{\left| {{{\widetilde h}_b}} \right|}^2} + W{N_0}}}{{{{\left| {{{\widetilde h}_b}} \right|}^2}}}\left( {{2^{\frac{{\overline R }}{W}}} - 1} \right).
\end{align}
\subsection{Power Region Characterization for semi-NOMA}
For semi-NOMA, the corresponding optimization problem is given by
\vspace{-0.6cm}
\begin{subequations}\label{S NOMA P}
\begin{align}
&\mathop {\min }\limits_{{W_m},{W_b},{p_s\ge0},{p_{b,m}\ge0},{p_{b,o}\ge0}} \;p_s+p_{b,m}+p_{b,o}\\
\label{S S P}{\rm{s.t.}}\;\;&\frac{{{W_m}I}}{{KL}}{\widetilde \varepsilon _K}\left( {\frac{{{p_s}{{\left| {{h_s}} \right|}^2}}}{{{W_m}{N_0}}}} \right) \ge \overline S ,\\
\label{SS S P}&{\widetilde \varepsilon _K}\left( {\frac{{{p_s}{{\left| {{h_s}} \right|}^2}}}{{{W_m}{N_0}}}} \right) \ge \overline \varepsilon  ,\\
\label{Rate S P}&{W_m}{\log _2}\left( {1 + \frac{{{p_{b,m}}{{\left| {{{\widetilde h}_b}} \right|}^2}}}{{{p_s}{{\left| {{{\widetilde h}_b}} \right|}^2} + {W_m}{N_0}}}} \right) + {W_b}{\log _2}\left( {1 + \frac{{{p_{b,o}}{{\left| {{h_b}} \right|}^2}}}{{{W_b}{N_0}}}} \right) \ge \overline R,\\
\label{S Bandwidth allocation Semi}&{W_m} + {W_b} = W,{W_m} \ge 0,{W_b} \ge 0.
\end{align}
\end{subequations}
\vspace{-1cm}

\noindent For any given bandwidth allocation, $\left( {{W_m},{W_b}} \right)$, the minimum required transmit power for the semantic transmission is $p_s^* \triangleq {p_K}\left( {\overline S,\overline \varepsilon ,{W_m}} \right)$, which is defined in \eqref{C3}. Then, problem \eqref{S NOMA P} is reduced to
\vspace{-0.6cm}
\begin{subequations}\label{S NOMA P 1}
\begin{align}
&\mathop {\min }\limits_{{p_{b,m}\ge0},{p_{b,o}\ge0}} \;p_{b,m}+p_{b,o}\\
\label{Rate S P 1}{\rm{s.t.}}&{W_m}{\log _2}\!\left(\! {1\! +\! \frac{{{p_{b,m}}{{\left| {{{\widetilde h}_b}} \right|}^2}}}{{{{p_K}\left( {\overline S,\overline \varepsilon ,{W_m}} \right)}{{\left| {{{\widetilde h}_b}} \right|}^2}\! + \!{W_m}{N_0}}}}\! \right)\! +\! \left( {W \!- \!{W_m}} \right){\log _2}\left(\! {1\! +\! \frac{{{p_{b,o}}{{\left| {{h_b}} \right|}^2}}}{{\left( {W\! - \!{W_m}} \right){N_0}}}} \!\right) \!\ge\! \overline R.
\end{align}
\end{subequations}
\vspace{-1cm}

\noindent It can be found that the optimal solution to problem \eqref{S NOMA P 1} also follows the ``water-filling'' power allocation scheme~\cite{Fundamentals}. We first ignore the non-negative constraint on ${p_{b,m}}$ and ${p_{b,o}}$. Then, the corresponding optimal ``water-filling'' power allocation solution is given by
\vspace{-0.2cm}
\begin{align}\label{ap}
{\widetilde p_{b,m}^*} = {\widetilde \lambda ^*}{W_m} - \frac{1}{{{{\widetilde h}_m}}},{\widetilde p_{b,o}^*} = {\widetilde \lambda ^*}\left( {W - {W_m}} \right) - \frac{1}{{{{\widetilde h}_b}}},
\end{align}
\vspace{-0.8cm}

\noindent where ${\widetilde h_m} \triangleq \frac{{{{\left| {{{\widetilde h}_b}} \right|}^2}}}{{{p_K}\left( {\overline S,\overline \varepsilon ,{W_m}} \right){{\left| {{{\widetilde h}_b}} \right|}^2} + {W_m}{N_0}}}$, ${\widetilde h_b} \triangleq \frac{{{{\left| {{h_b}} \right|}^2}}}{{{\left( {W - {W_m}} \right)}{N_0}}}$, and ${\widetilde \lambda ^*} \triangleq \frac{{{2^{\frac{{\overline R }}{W}}}}}{{{{\left( {{W_m}{{\widetilde h}_m}} \right)}^{\frac{{{W_m}}}{W}}}{{\left( {{\left( {W - {W_m}} \right)}{{\widetilde h}_b}} \right)}^{\frac{{{\left( {W - {W_m}} \right)}}}{W}}}}}$.
Therefore, the optimal solution to problem \eqref{S NOMA P 1} can be expressed as
\vspace{-0.2cm}
\begin{align}
\left( {p_{b,m}^*,p_{b,o}^*} \right) = \left\{ \begin{gathered}
  \left( {0,\widehat p_{b,o}^*} \right),\;\;{\rm{if}}\;\;\widetilde p_{b,m}^* \le 0,\; \hfill \\
  \left( {\widehat p_{b,m}^*,0} \right),\;\;{\rm{if}}\;\;\widetilde p_{b,o}^* \le 0, \hfill \\
  \left( {\widetilde p_{b,m}^*,\widetilde p_{b,o}^*} \right),\;{\rm{otherwise}}, \hfill \\
\end{gathered}  \right.
\end{align}
\vspace{-0.4cm}

\noindent where $\widehat p_{b,o}^* \triangleq \frac{1}{{{{\widetilde h}_b}}}{2^{\frac{{\overline R }}{{W - {W_m}}}}}$ and $\widehat p_{b,m}^* \triangleq \frac{1}{{{{\widetilde h}_m}}}{2^{\frac{{\overline R }}{{{W_m}}}}}$. As a result, by employing the one-dimensional search over ${{W_m} \in \left[ {0,W} \right]}$, the optimal solution to problem \eqref{S NOMA P} can be obtained as follows:
\vspace{-0.4cm}
\begin{align}
P_{\min }^{\rm{S-N}}\left( {\overline S,\overline \varepsilon  ,\overline R } \right) =\mathop {\arg \min }\limits_{{W_m} \in \left[ {0,W} \right]} \;\left( {{p_K}\left( {\bar S,\bar \varepsilon ,{W_m}} \right) + p_{b,m}^* + p_{b,o}^*} \right).
\end{align}
\vspace{-1cm}

\noindent Since OMA and NOMA can be regarded as special cases of semi-NOMA, we have $P_{\min }^{\rm{S-N}}\left( {\overline S,\overline \varepsilon  ,\overline R } \right)$\\$ \le P_{\min }^{\rm{O}}\left( {\overline S,\overline \varepsilon  ,\overline R } \right)$ and $P_{\min }^{\rm{S-N}}\left( {\overline S,\overline \varepsilon  ,\overline R } \right) \le P_{\min }^{\rm{N}}\left( {\overline S,\overline \varepsilon  ,\overline R } \right)$.
\section{Numerical Examples}
In this section, we provide numerical examples to evaluate the performance of the three proposed MA schemes to facilitate the heterogeneous semantic and bit transmission. Let $d_s$ and $d_b$ denote the AP-S-user distance and the AP-B-user distance in meters, respectively. The small-scale fading of the AP-S/B-user links are modelled as Rayleigh fading channels, which are generated as independent circularly symmetric complex Gaussian random variables with zero mean and unit variance. The distance-dependent path loss of the corresponding link is modelled as $\rho  = {\rho _0}{\left( {{1 \mathord{\left/
 {\vphantom {1 d}} \right.
 \kern-\nulldelimiterspace} d}} \right)^\beta }$, where ${\rho _0}=-30$ dB denotes the reference path loss at 1 meter, $\beta =4$ denotes the path loss exponent, and $d$ denotes the corresponding link distance in meters. The total available frequency bandwidth is set to $W = 1$ MHz. The power spectral density of the received white Gaussian noise is ${N_0} =  - 140$ dBm/Hz, i.e., an average noise power of $- 80$ dBm over the bandwidth of 1 MHz.
\subsection{SvB Rate Region Comparison}
In this subsection, we first investigate the SvB rate region achieved by the three proposed MA schemes in the power-sufficient case.
\subsubsection{Impact of Users' Channels Differences} 
\begin{figure}[!t]
\centering
\begin{minipage}[t]{0.48\linewidth}
\includegraphics[width=3in]{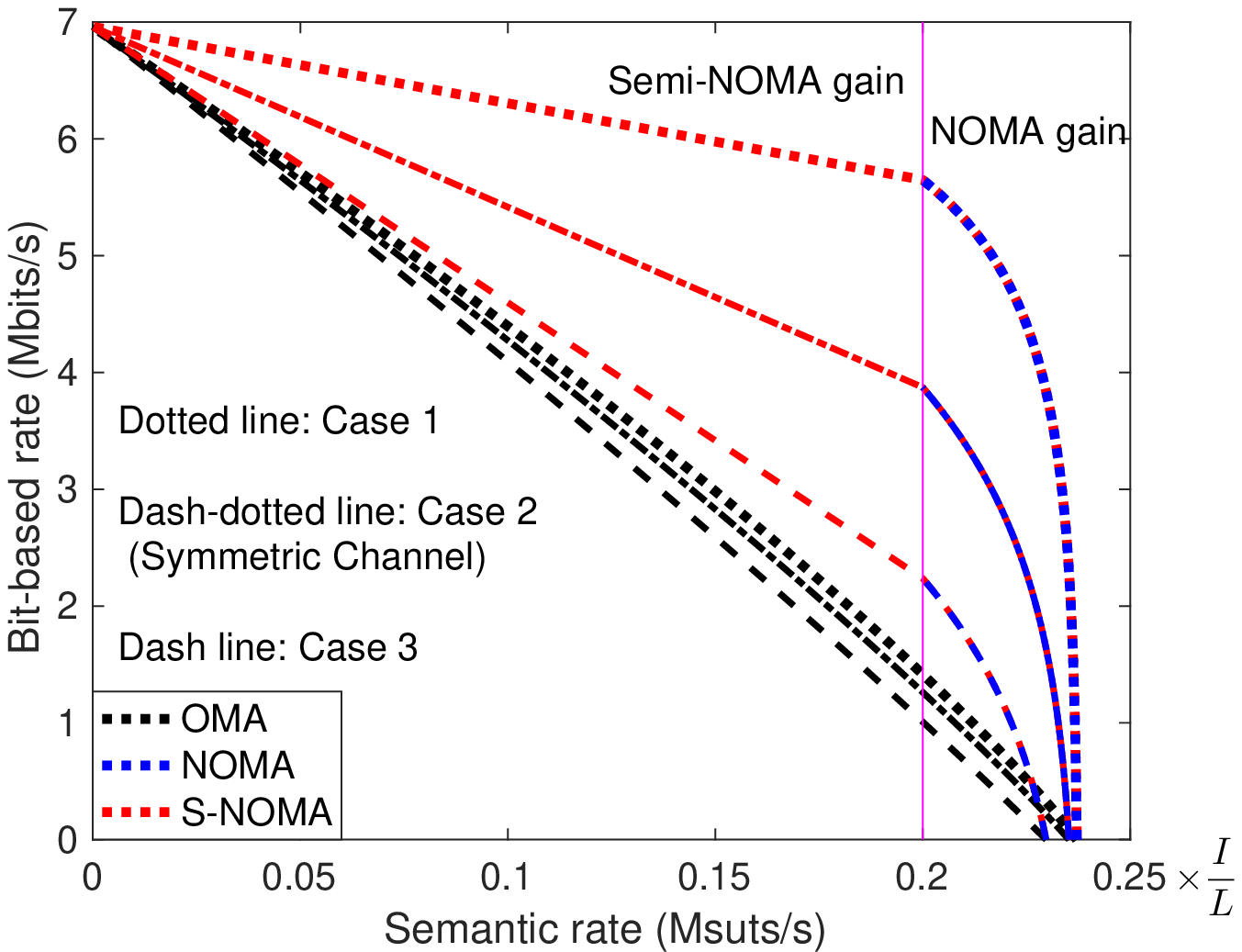}\\
\caption{SvB rate region comparison of each proposed MA scheme under three channel difference cases.}\label{SvB}
\end{minipage}
\quad
\begin{minipage}[t]{0.48\linewidth}
\includegraphics[width=3in]{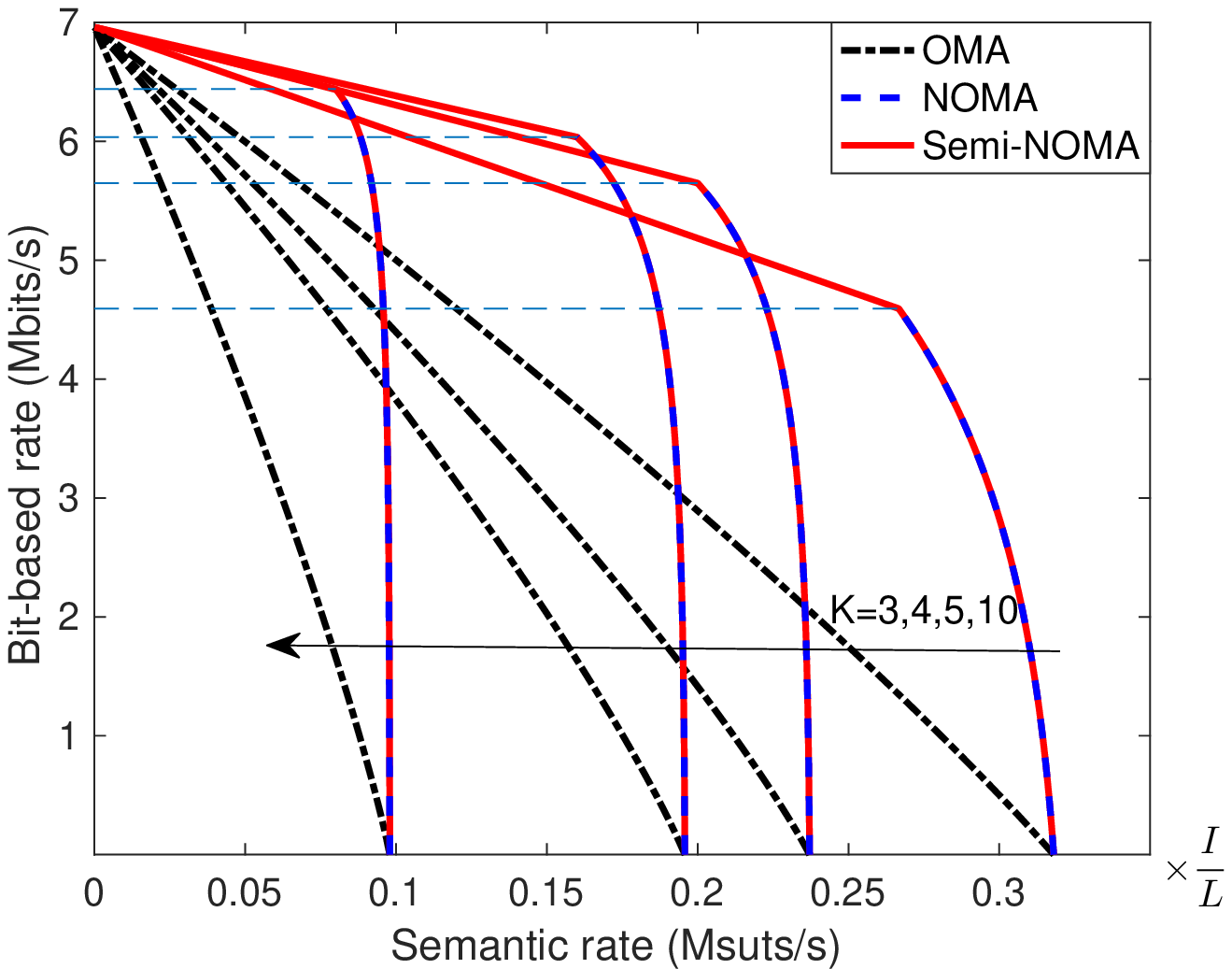}\\
\caption{SvB rate region comparison of each proposed MA scheme under different $K$.}\label{SvBvK}
\end{minipage}
\end{figure}

In Fig. \ref{SvB}, we first study the impact of the channel difference between the S- and B-users on the achieved SvB rate region for each proposed MA scheme. In particular, we consider the following three cases: (1) \textbf{Case 1}: we consider the case when the channel gain of the S-user is larger than that of the B-user, and set ${d_s} = 20$ meter and ${d_b} = 30$ meter; (2) \textbf{Case 2 (symmetric channel)}: we consider a special case when the channel gains of the S- and B-users are the same, and set ${d_s} = 30$ meter and ${d_b} = 30$ meter; (3) \textbf{Case 3}: we consider the case when the channel gain of the S-user is smaller than that of the B-user, and set ${d_s} = 40$ meter and ${d_b} = 30$ meter. For other parameters, we set $P=30$ dBm, $\overline \varepsilon  =0.8$, and $K=4$ in each case. As it can be observed from Fig. \ref{SvB}, the SvB rate region achieved by semi-NOMA always strictly contains those achieved by OMA and NOMA in each case. This is consistent with \textbf{Remark 2} since semi-NOMA not only specializes to OMA and NOMA but also provides more flexible transmission options. This result validates the effectiveness of the proposed semi-NOMA scheme to facilitate the heterogeneous semantic and bit transmission. It can also be observed that, compared to OMA and semi-NOMA, where the bandwidth allocated to the semantic stream can be adjusted, the SvB rate region achieved by NOMA is quite restricted. This is because the frequency band is fully shared in NOMA, which always requires a considerable transmit power to be allocated to the semantic stream to satisfy the minimum semantic similarity constraint. Moreover, it can be found that in the high semantic rate regime, the performance of semi-NOMA and NOMA is the same and outperforms OMA. This implies that to achieve a high semantic rate, NOMA is still an effective MA scheme. Interestingly, it can be seen that OMA is always strictly suboptimal even if in the symmetric channel case. This observation is different from the conventional bit-based transmission, where OMA can achieve the same performance as NOMA for the symmetric channel. This result underscores the importance of employing a non-orthogonal-type MA scheme for the heterogeneous semantic and bit transmission. By comparing the achieved SvB rate region in the three cases, it can be observed that, for a fixed B-user's channel gain, the SvB rate region is enlarged when the S-user's channel gain increases. In particular, the corresponding region enlargement in semi-NOMA is more pronounced than that in OMA. This also confirms the effectiveness of the proposed semi-NOMA scheme and reveals a useful guidance, namely, that it is preferable to pair an S-user having a higher channel gain and a B-user having a lower channel gain together to achieve a higher performance when employing semi-NOMA.
\subsubsection{Impact of $K$} In Fig. \ref{SvBvK}, we continue to study the impact of $K$ on the achieved SvB rate region, where $P=30$ dBm, $\overline \varepsilon  =0.8$, ${d_s} = 20$ meter, and ${d_b} = 30$ meter. It can be observed that when $K$ increases, the maximum achieved semantic rate decreases. This is because a larger $K$ means that more semantic symbols have to be transmitted for delivering one word in the sentence, which decreases the corresponding semantic rate, see \eqref{effective SSE}. For NOMA, it can also be observed that although a smaller $K$ leads to a higher maximum semantic rate, the corresponding achieved SvB rate region does not contain those regions achieved by larger $K$. The reason behind this can be explained as follows. As it can be observed from Fig. \ref{SS_result}, to achieve the same semantic similarity, a larger $K$ requires a lower SNR, i.e., less transmit power could be allocated to the semantic stream in NOMA. As a result, to achieve the boundary point $\left( {{S_{\min }^{\rm{N}}},R_{\max }^{\rm{N}}} \right)$ in NOMA, a larger $K$ can achieve a higher $R_{\max }^{\rm{N}}$. For semi-NOMA, when $K=3, 4, 5$, the SvB rate region achieved by a smaller $K$ also does not strictly contain that achieved by a larger $K$. However, when $K$ increases to $10$, the corresponding achieved SvB rate region is contained in that achieved by $K=4$. This reveals that increasing $K$ would be ineffective in performance improvement for semi-NOMA. However, in contrast to semi-NOMA and NOMA, it can be observed that for OMA, the SvB rate region achieved by smaller $K$ strictly contains those achieved by larger $K$. Moreover, it can also be observed that for each $K$, the semi-NOMA SvB rate region always contains those achieved by OMA and NOMA. This also confirms the effectiveness of the proposed semi-NOMA scheme.
\subsection{Power Region Comparison}
In this subsection, we investigate the power region of each proposed MA scheme. We set ${d_s} = 20$ meter and ${d_b} = 30$ meter. All the following results were obtained by averaging over 5000 channel realizations.
\subsubsection{Impact of $\overline S$} 
\begin{figure}[!t]
\centering
\begin{minipage}[t]{0.48\linewidth}
\includegraphics[width=3in]{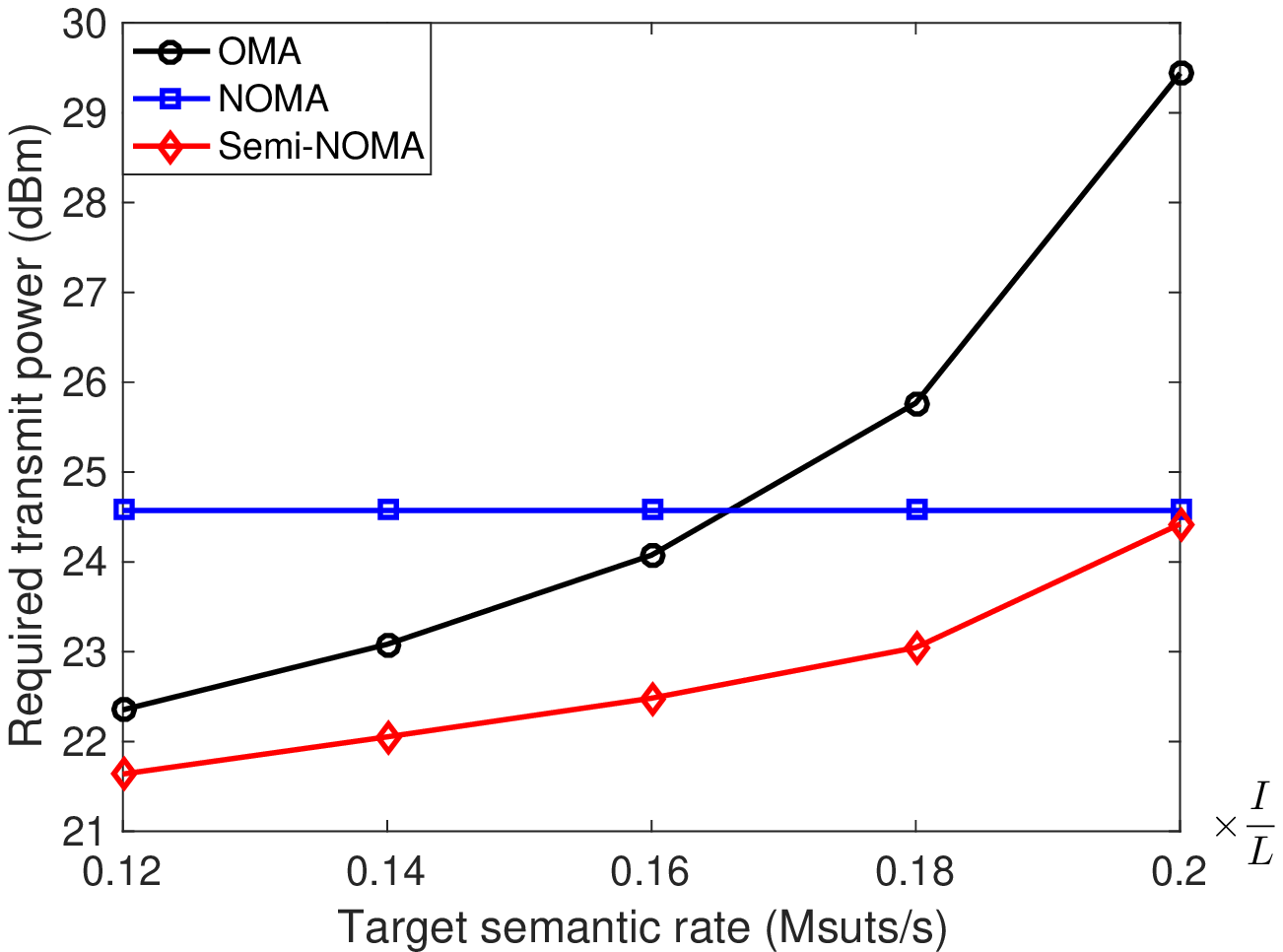}\\
\caption{Required minimum transmit power versus $\overline S$ for $K=4$, $\overline \varepsilon  =0.8$, and $\overline R=0.8$ Mbits/s.}\label{PvS}
\end{minipage}
\quad
\begin{minipage}[t]{0.48\linewidth}
    \includegraphics[width=3in]{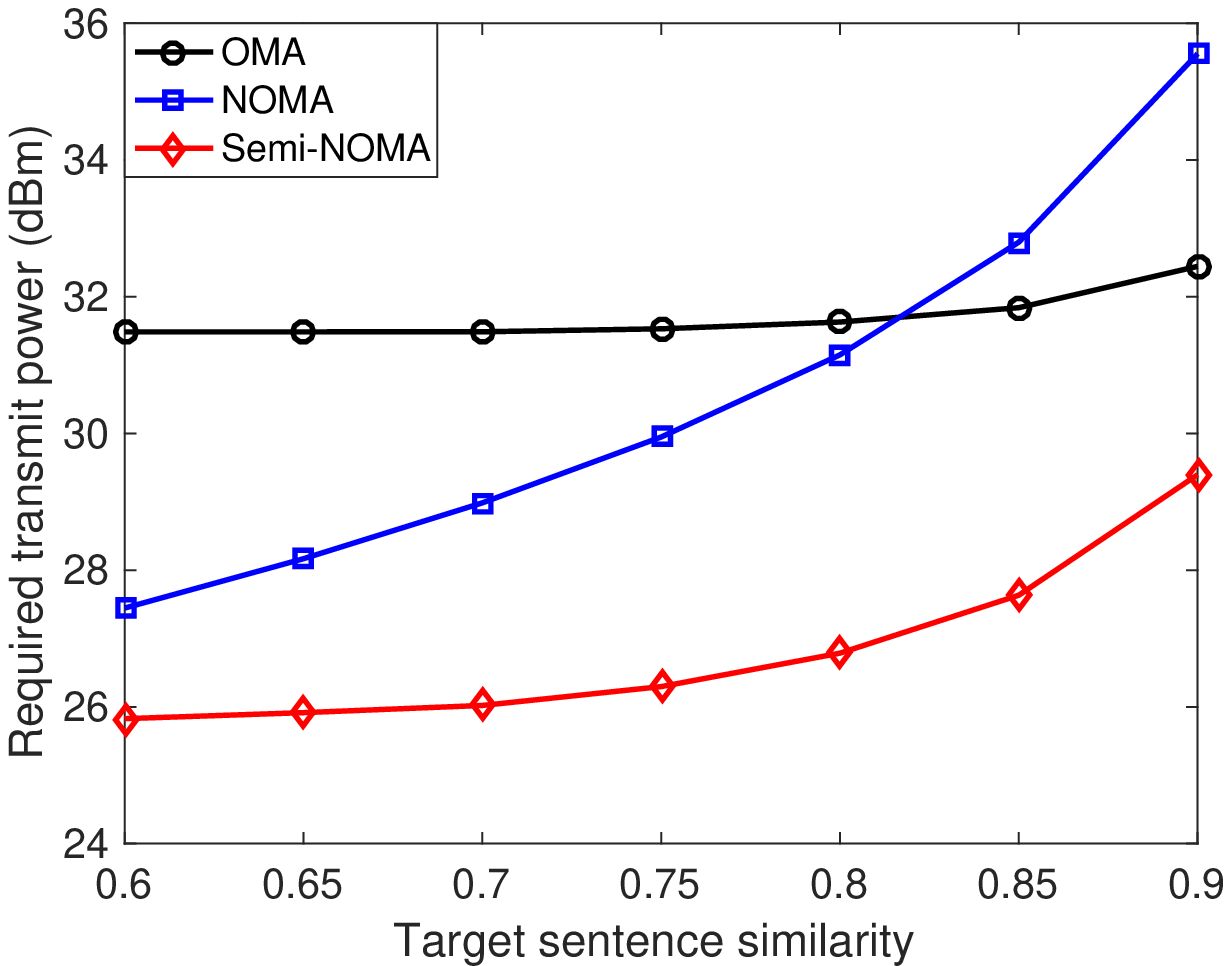}\\
\caption{Required minimum transmit power versus $\overline \varepsilon$ for $K=4$, $\overline S  =0.15\frac{I}{L}$ Msuts/s, and $\overline R=2$ Mbits/s.}\label{PvSS}
\end{minipage}
\end{figure}

In Fig. \ref{PvS}, we investigate the required minimum transmit power versus the target semantic rate, $\overline S$. We set $K=4$, $\overline \varepsilon  =0.8$, and $\overline R=0.8$ Mbits/s. As can be seen from Fig. \ref{PvS}, the required transmit power of semi-NOMA and OMA increases as $\overline S$ increases. This is expected since a higher $\overline S$ requires a larger bandwidth and/or a higher semantic similarity, thus increasing the power consumption. However, for NOMA, the power consumption remains unchanged as $\overline S$ increases. This is because the required semantic similarity constraint causes the semantic rate to be achieved automatically in NOMA. As a result, the minimum transmit power depends on the semantic similarity constraint, instead of the target semantic rate. Different from the conventional bit-based communication, where NOMA can achieve no worse performance than OMA, NOMA and OMA in the heterogeneous semantic and bit transmission are superior in the high and low $\overline S$ regimes, respectively. This is because in the low $\overline S$ regime, the adjustment of bandwidth allocation in OMA helps to reduce the required transmit power to achieve $\overline S$, which is not possible for NOMA. In the high $\overline S$ regime, as a large bandwidth should be allocated to the semantic stream, the employment of orthogonal sub-bands in OMA leads to only limited bandwidth available for the bit transmission, thus consuming a sufficiently high power to achieve $\overline R$. However, in this case, the spectrum sharing in NOMA helps to reduce the power consumption to achieve $\overline R$. As the proposed semi-NOMA scheme provides flexible resource allocation, it is expected that it achieves the best performance among all schemes.
\subsubsection{Impact of $\overline \varepsilon$} In Fig. \ref{PvSS}, we study the required minimum transmit power versus the target semantic similarity, $\overline \varepsilon$. We set $K=4$, $\overline S  =0.15\frac{I}{L}$ Msuts/s, and $\overline R=2$ Mbits/s. As can be seen, the required transmit power of all schemes increases as $\overline \varepsilon$ increases since a higher received SNR needs to be achieved. It can also be observed that the required transmit power in OMA and semi-NOMA is less sensitive to $\overline \varepsilon$ than that in NOMA. This is because in NOMA, the frequency band is fully occupied by the semantic stream (i.e., a higher received noise power), and more transmit power is required to achieve the target $\overline \varepsilon$. This underscores the importance of bandwidth allocation in the considered heterogeneous semantic and bit transmission.
\subsubsection{Impact of $K$} 
\begin{figure}[!t]
  \centering
  \includegraphics[width=3in]{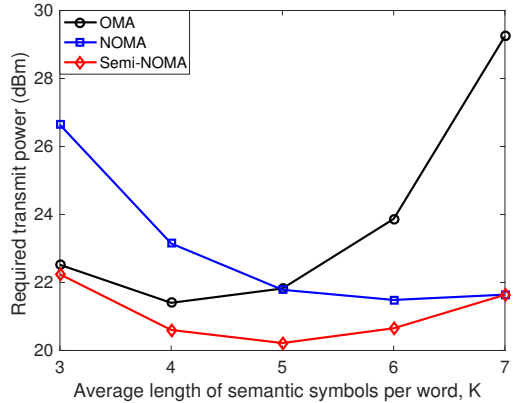}\\
  \caption{Required minimum transmit power versus $K$ for $\overline S  =0.12\frac{I}{L}$ Msuts/s, $\overline \varepsilon  =0.8$, and $\overline R=0.8$ Mbits/s.}\label{PvK}
\end{figure}

In Fig. \ref{PvK}, we investigate the required minimum transmit power versus the semantic encoding/decoding scheme, $K$. We set $\overline S  =0.12\frac{I}{L}$ Msuts/s, $\overline \varepsilon  =0.8$, and $\overline R=0.8$ Mbits/s. It can be observed that as $K$ increases, the required transmit power of all schemes in general first decreases and then increases after a specific $K$. For the considered scenario, $K=5$, $K=6$, and $K=4$ are the optimal semantic encoding/decoding scheme for semi-NOMA, NOMA, and OMA, respectively. The reasons behind this trend can be explained as follows. For NOMA under the given $\overline S$, $\overline \varepsilon$, and $\overline R$ requirements, $\overline \varepsilon$ dominates the power consumption when $3 \le K \le 6$ due to the employment of full frequency band for the semantic transmission. As it can be seen from Fig. \ref{SS_result}, a smaller $K$ leads to a higher transmit power being allocated to the semantic stream to achieve the same semantic similarity. This also increases the transmit power required to achieve $\overline R$ due to the inter-stream interference in NOMA. Therefore, increasing $K$ will reduce the power consumption. However, when $K$ increases to 7, $\overline S$ becomes dominated in power consumption, thus leading to a higher transmit power to achieve a higher semantic similarity compared to the case of $K=6$. For OMA, it also follows that more transmit power has to be allocated to achieve $\overline \varepsilon$ when $K=3$ as compared to $K=4$\footnote{This result is different from that presented in Fig. \ref{SvBvK}, where the SvB rate region achieved by smaller $K$ always contains that achieved by larger $K$. This is because for the SvB rate region presented in Fig. \ref{SvBvK}, we only focus on the power-sufficient case, while both the power-sufficient and power-limited cases may occur in the power region characterization.}, see Fig. \ref{SS_result}. However, when $K$ further increases, considerable bandwidth should be allocated to the semantic transmission to achieve $\overline S$. This, in turn, significantly reduces the bandwidth that can be allocated to the bit transmission, which greatly increases the power consumption to achieve $\overline R$. For semi-NOMA with $3 \le K \le 5$ also follows the principle that a smaller $K$ requires more transmit power to achieve the same $\overline \varepsilon$ in the semantic transmission. Although $K=5$ may require considerable bandwidth to be allocated to the semantic stream, the partial spectrum sharing mechanism in semi-NOMA helps to reduce the power consumption to achieve $\overline R$. However, when $K$ further increases (i.e., $K=6$ and 7), almost all the bandwidth should be allocated to the shared frequency sub-band to achieve $\overline S$. In this case, a small orthogonal bit-only frequency sub-band is available and the bit transmission mainly relies on the shared frequency sub-band, which increases the power consumption to achieve $\overline R$. The above results also confirm the effectiveness of the proposed semi-NOMA given its flexible resource allocation feature, which improves the resource efficiency. It also opens up an interesting research direction for the optimization of the semantic encoding/decoding scheme of semantic transmission since it can be observed that the achieved performance is sensitive to $K$. 
\section{Conclusions and Future Work}
The new heterogeneous semantic and bit transmission has been investigated in this paper with a particular focus on the MA design. The semantic rate was adopted for measuring the performance of the semantic transmission. To solve the problem of the lack of a closed-form expression for the semantic similarity, the data regression method was proposed by approximating it with a generalized logistic function. Based on this approximation, a heterogeneous semantic and bit communication framework was proposed, where an AP simultaneously sends the semantic and bit streams to the S- and B-users. To facilitate this heterogeneous transmission, three MA schemes were presented and compared, namely OMA, NOMA, and semi-NOMA. The SvB rate region and the power region achieved by the three proposed MA schemes were characterized. In contrast to the conventional bit-based transmission, our results revealed that NOMA does not necessarily outperform OMA, and semi-NOMA is promising given its flexible transmission strategies. The presented numerical examples validated the analysis and showed the superiority of semi-NOMA over other two MA schemes. They also revealed a useful guidance that it is preferable to pair an S-user with a higher channel gain with a B-user with a lower channel gain for maximizing the performance gain of semi-NOMA. Moreover, our results also showed that the optimization of the semantic encoding/decoding scheme is important for improving the semantic transmission performance. 

This paper puts forward an interesting research direction of designing the heterogeneous semantic and bit transmission towards next-generation wireless networks, and there are numerous open research problems, some of which are briefly described to motivate future work.
\begin{itemize}
  \item \emph{Heterogeneous transmission design with multiple S- and B-users:} This paper considered a fundamental model with one S-user and one B-user. In practice, there may exist multiple S- and B-users that have to be served by the AP. In such a general case, one potential solution is to first divide all users into several clusters, which consist of one S-user and one B-user. The AP uses orthogonal frequency-division multiple access (OFDMA) to serve each cluster by assigning orthogonal subcarriers, while the proposed semi-NOMA can be employed within each cluster. How to jointly design the user pairing and resource allocation (between each cluster as well as within each cluster) is an interesting but challenging problem, which requires further research. 
  \item \emph{Performance limit characterization of the heterogeneous transmission with multimodal data:} The semantic rate considered in this paper is used for characterizing the performance of text semantic communications. When semantic communications transmit multimodal data (e.g., image, video, and speech), the corresponding fundamental performance limit characterization problem in the heterogeneous transmission is another open problem. To address this issue, a unified performance metric for evaluating the performance of multimodal semantic communications may need to be developed, which merits further investigation.
  \item \emph{Opportunistic semantic and bit transmission design:} This paper assumed that each user can only use one specific transmission method, i.e., semantic or bit-based transmission. In the current literature~\cite{DeepSC_T}, semantic communications are shown to be superior to conventional bit communications mainly in the low/moderate SNR regime. Given the randomness of users' channel conditions, one promising approach is to develop opportunistic semantic and bit transmission policy for fully reaping the benefits of both semantic and bit transmission. This also constitutes an interesting and challenging new research problem.
\end{itemize}

\bibliographystyle{IEEEtran}
\bibliography{mybib}
\end{document}